\documentclass[aps,prd,twocolumn,groupedaddress]{revtex4-2}

\usepackage{float} 
\usepackage{graphicx}
\def\rf#1{(\ref{eq:#1})}
\def\lab#1{\label{eq:#1}}

\def\br{\begin{eqnarray}}
\def\er{\end{eqnarray}}
\def\be{\begin{equation}}
\def\ee{\end{equation}}

\def\({\left(}
\def\){\right)}

\relax

\def\trace{\widehat{\rm Tr}}

\newcommand\sbr[2]{\left\lbrack\,{#1}\, ,\,{#2}\,\right\rbrack}

\def\tp0{\Theta_{+}^{(0)}}
\def\tm0{\Theta_{-}^{(0)}}

\def\ve{\varepsilon}

\def\u2{\mid u\mid^2}

\def\z2{\mid z\mid^2}

\def\w2{\mid w\mid^2}

\def\f#1#2#3 {f^{#1#2}_{#3}}

\def\win1{{\sf w_{1+\infty}}}

\def\Win1{{\sf W_{1+\infty}}}

\def\rlx{\relax\leavevmode}
\def\inbar{\vrule height1.5ex width.4pt depth0pt}
\def\IZ{\rlx\hbox{\sf Z\kern-.4em Z}}
\def\IR{\rlx\hbox{\rm I\kern-.18em R}}
\def\IC{\rlx\hbox{\,$\inbar\kern-.3em{\rm C}$}}
\def\IN{\rlx\hbox{\rm I\kern-.18em N}}
\def\IO{\rlx\hbox{\,$\inbar\kern-.3em{\rm O}$}}
\def\IP{\rlx\hbox{\rm I\kern-.18em P}}
\def\IQ{\rlx\hbox{\,$\inbar\kern-.3em{\rm Q}$}}
\def\IF{\rlx\hbox{\rm I\kern-.18em F}}
\def\IG{\rlx\hbox{\,$\inbar\kern-.3em{\rm G}$}}
\def\IH{\rlx\hbox{\rm I\kern-.18em H}}
\def\II{\rlx\hbox{\rm I\kern-.18em I}}
\def\IK{\rlx\hbox{\rm I\kern-.18em K}}
\def\IL{\rlx\hbox{\rm I\kern-.18em L}}
\def\one{\hbox{{1}\kern-.25em\hbox{l}}}
\def\0#1{\relax\ifmmode\mathaccent"7017{#1}%
B        \else\accent23#1\relax\fi}

\begin{document}


\title{A Generalised Self-Duality for the Yang-Mills-Higgs System}


\author{L. A. Ferreira and H. Malavazzi}
\email[]{laf@ifsc.usp.br, henrique.malavazzi@usp.br}
\affiliation{Instituto de F\'\i sica de S\~ao Carlos; IFSC/USP;
Universidade de S\~ao Paulo, USP\\
Caixa Postal 369, CEP 13560-970, S\~ao Carlos-SP, Brazil}


\date{\today}

\begin{abstract}
Self-duality is a very important concept in the study and applications of topological solitons in many areas of Physics. The rich mathematical structures underlying it lead, in many cases, to the development of exact and non-perturbative methods. We present a generalization of the Yang-Mills-Higgs system by the introduction of scalar fields assembled in a symmetric and invertible matrix $h$ of the same dimension as the gauge group. The coupling of such new fields to the gauge and Higgs fields is made by replacing the Killing form, in the contraction of the group indices, by the matrix $h$ in the kinetic term for the gauge fields, and by its inverse in the Higgs field kinetic term. The theory is conformally invariant in the three dimensional space $\IR^3$. An important aspect of the model is that for practically all configurations of the gauge and Higgs fields the new scalar fields adjust themselves to solve the modified self-duality equations. We construct solutions using a spherically symmetric ans\"atz and show that the 't Hooft-Polyakov monopole becomes a self-dual solution of such modified Yang-Mills-Higgs system. We use an ans\"atz based on the conformal symmetry to construct vacuum solutions presenting non-trivial toroidal magnetic fields.

\end{abstract}


\maketitle



\section{Introduction}
\label{sec:intro}
\setcounter{equation}{0}

Topological solitons play a fundamental role in the study of non-linear phenomena in many areas of science. Their stability, inherited from non-trivial topological structures, makes them ideal candidates to describe  excitations in some sectors of the theory, specially strong coupling regimes. Examples of topological solitons range from kinks in $(1+1)$-dimensions, to vortices and magnetic Skyrmions in $(2+1)$-dimensions, magnetic monopoles and Skyrmions in $(3+1)$-dimensions, and instantons in four dimensional Euclidean spaces. They find applications from high energy physics to condensed matter physics and in non-linear phenomena in general \cite{mantonbook,shnirbook,shnirbookmonopoles}. 

There is a class of topological solitons however, that deserves a special attention as they reveal deeper mathematical structures in the theory, which may lead to the development of some exact and non-perturbative methods. They present two main properties: first, they are classical solutions of the so-called self-duality equations which are first order differential equations that imply the second order Euler-Lagrange equations of the theory. Second, on each topological sector of the theory there is a lower bound on the static energy, or Euclidean action, and the self-dual solitons saturate that bound. Therefore, self-dual solitons are very stable.

The fact that one has to perform one integration less to construct self-dual solitons, as compared to the usual topological solitons, is not linked to the use of any dynamically conserved quantity. In all known examples, the relevant topological charge admits an integral representation, and so there exists a density of topological charge. As such charge is invariant under any smooth (homotopic) variations of the fields, it leads to local identities, in the form of second order differential equations, that are satisfied by any regular configuration of the fields, not necessarily solutions of the theory. The magic is that such identities become the Euler-Lagrange equations of the theory when the self-duality equations are imposed. That may happen even in the cases where there is no lower bound on the energy or Euclidean action. 

By exploring such ideas it was possible to develop the concept of generalized self-dualities where one can construct, from one single topological charge, a large class of field theories possessing self-dual sectors \cite{genbps}. In $(1+1)$-dimensions it was possible to construct field theories, with any number of scalar fields, possessing self-dual solitons, and so generalizing what is well known in theories with one single scalar field, like sine-Gordon and $\lambda\, \phi^4$ models \cite{sd2d,bpscomments}. In addition, exact self-dual sectors were constructed for Skyrme type theories by the addition of extra scalar fields \cite{bpswojtek,bpsshnir,laf2017,us}, and concrete applications have been made to nuclear matter \cite{nuclear}.

In this paper we apply such ideas and methods to the Yang-Mills-Higgs system in $(3+1)$-dimensions. In this case, the relevant topological charge is the magnetic charge defined by the integral 
\be
\int_{\IR^3} d^3x\,\ve_{ijk}{\rm Tr}\(F_{ij}\,D_k\Phi\)
\lab{topchargeymhintro}
\ee
where $F_{ij}=\partial_iA_j-\partial_jA_i+i\,e\,\sbr{A_i}{A_j}$, is the field tensor, $A_i=A_i^a\,T_a$, the gauge field, and $\Phi=\Phi_a\,T_a$, the Higgs field in the adjoint representation of a simple, compact, Lie group $G$, with generators $T_a$, $a=1,2,\ldots {\rm dim}\,G$. In addition, $D_i *=\partial_i*+i\,e\,\sbr{A_i}{*}$ is the covariant derivative in the adjoint representation of $G$. 

The generalized self-duality equations are given by
\be
\frac{1}{2}\,\ve_{ijk}\,F_{jk}^b\,h_{ba}=\pm \(D_i\Phi\)^a
\lab{sdeqsymhintro}
\ee
where $h_{ab}$, $a\,,\, b=1,2,\ldots {\rm dim}\,G$, is a symmetric invertible matrix of scalar fields. Following \cite{genbps}, we show in section \ref{sec:sd}, the identities following from the invariance of \rf{topchargeymhintro}, under smooth variations of the fields, combined with the self-duality equations \rf{sdeqsymhintro}, imply the Euler-Lagrange equations associated to the static energy functional given by 
\be
E_{YMH}=\int d^3x\,\left[\frac{1}{4}\,h_{ab}\,F_{ij}^a\,F_{ij}^b+\frac{1}{2}\,h^{-1}_{ab}\,\(D_i\Phi\)^a\,\(D_i\Phi\)^b\right]
\lab{energyymhintro}
\ee
In fact, they imply not only the Euler-Lagrange equations associated to the gauge and Higgs fields, but also the ones associated to the scalar fields $h_{ab}$. 

Clearly, in the case where the matrix $h$ is the unit matrix the self-duality equations \rf{sdeqsymhintro} becomes the usual Bogomolny equations \cite{bogo}, and \rf{energyymhintro} becomes the static energy functional for the Yang-Mills-Higgs system in the Prasad-Sommerfield limit \cite{prasad}. Modifications of the Yang-Mills-Higgs system have been considered in \cite{dahora,dahora2,bazeia1,bazeia2,zhang} where the kinetic terms of gauge and Higgs fields are multiplied by functionals of the modulus of the Higgs field, without the introduction of new  fields. 

The introduction of the scalar fields $h_{ab}$ brings in some novel features. They make the static sector of the theory conformally invariant in the three dimensional space $\IR^3$, and that plays an important role in many aspects of the theory, specially in the construction of solutions. The eigenvalues of the matrix $h$ have to be positive to make the energy \rf{energyymhintro} positive definite. That is guaranteed in most of the cases, but as we will show, it is possible to use the conformal symmetry to build an ans\"atz to construct vacuum solutions, with vanishing energy and topological charge, and presenting non-trivial magnetic fields in toroidal configurations. We give an example where the toroidal magnetic field possesses a new non-trivial topological charge resembling the concept of helicity used in magnetohydrodynamics. Clearly, for such non-trivial vacuum configurations the eigenvalues of $h$ are not all positive, and it would be interesting to investigate their stability. 

The scalar fields $h_{ab}$ transform under the symmetric part of the tensor product of the adjoint representation of the gauge group with itself. Their asymptotic value at spatial infinity may be related to some pattern of spontaneous symmetry breaking. Note, that we do not have a Higgs potential in \rf{energyymhintro}, neither are considering the Prasad-Sommerfield limit of it. As an example, we consider the usual spherically symmetric 't Hooft-Polyakov ans\"atz for the case $G=SU(2)$, and show that for any configuration in such an ansatz, two of the three eigenvalues of $h$ are equal,  pointing to some spontaneous breaking of the symmetry to $U(1)$. Indeed,  some configurations behave at spatial infinity such that two eigenvalues go to unity and the third to zero, leaving $h$ invariant under a $U(1)$ subgroup. 

Finally, the introduction of the scalar fields $h_{ab}$ enlarge the space of solutions considerably. A special role is played by the matrices $\tau_{ab}\equiv \frac{1}{2}\, F_{ij}^a\,F_{ij}^b$, and $\sigma_{ab}\equiv-\frac{1}{2}\,\ve_{ijk}\,F_{ij}^a\,\(D_k\Phi\)^b$. For the configurations of the gauge fields such that the matrix $\tau$ is invertible, one can show that the matrix $h$ given by $h=\pm\, \tau^{-1}\,\sigma$, solves the self-duality equations \rf{sdeqsymhintro}. Therefore, the scalar fields act as spectators adjusting themselves to the gauge and Higgs fields configurations, and solving the self-duality equations. In the cases where $\tau$ is singular it seems that some components of $h$ get undetermined but still one gets a solution for such configurations. In fact, that happens in one of our examples of vacuum configurations with non-trivial toroidal magnetic fields. So, there is still a lot to be understood about the physical role of the scalar fields $h_{ab}$. We comment however,  that if one diagonalizes it, i.e. $h=M\,h_D\,M^T$, with $M$ being an orthogonal matrix and $h_D$ diagonal, the fields in $h_D$ can be interpreted as dilaton fields leading to the conformal symmetry of the theory in the three dimensional space $\IR^3$. The $M$ fields relate, in many cases, to the Wilson line operator in the adjoint representation and lead to dressed quantities, namely field tensor and covariant derivative of the Higgs field,  that become gauge invariant. 

The paper is organized as follows. In section \ref{sec:sd} we present the ideas about the generalized self-duality and its features. In section \ref{sec:ymh} we discuss the properties of our modified Yang-Mills-Higgs system, construct the generalized self-duality equations and discuss their consequences. In section \ref{sec:su2} we use the well known 't Hooft-Polyakov spherically symmetric ans\"atz for the gauge group $G=SU(2)$, and construct new magnetic monopoles solutions. We show that the usual 't Hooft-Polyakov magnetic monopole becomes a self-dual solution of our modified Yang-Mills-Higgs system, even in the absence of a Higgs potential. In section \ref{sec:toroidal} we use the conformal symmetry to build an ans\"atz and construct new solutions for our theory. The subtlety here is that there seems to be no regular solutions with non-trivial energy and topological charge. We are able however, to construct vacuum solutions, with vanishing energy and topological charge, but with non-trivial toroidal magnetic field configurations. In one of the examples, the solution presents a new non-trivial topological charge similar to the concept of helicity used in magnetohydrodynamics. Then in section \ref{sec:conclusions} we present our conclusions, and in the appendix \ref{app:conformal} we show that the modified Yang-Mills-Higgs system is conformally invariant in the three dimensional space $\IR^3$.

\section{Generalised Self-Duality}
\label{sec:sd}
\setcounter{equation}{0}

The concept of self-duality has been used in Physics and Mathematics for a long time and in several contexts \cite{bogo,prasad,belavincp1,belavininstanton}. Basically, the self-duality equations are in general first order differential equations such that their solutions also solve the second order (static) Euler-Lagrange (EL) equations. In addition, those solutions saturate a bound on the static energy, or Euclidean action, related to a topological charge. The fact that the solutions are constructed by performing one integration less than what the EL equations would require, is not a consequence of the use of dynamical conservation laws. As explained in \cite{genbps}, it is related to the existence of a topological invariant that possesses an integral representation. Indeed, consider a field theory that possesses a topological charge with a integral representation of the form
\be
Q=\int d^dx\, {\cal A}_{\alpha}\,{\widetilde{\cal A}}_{\alpha}
\lab{topcharge}
\ee
where  ${\cal A}_{\alpha}$ and ${\widetilde{\cal A}}_{\alpha}$ are functionals of the fields of the theory and their first derivatives only, and where the index $\alpha$ stands for any type of indices, like vector, spinor, internal, etc, or groups of them. The fact that $Q$ is topological means that it is invariant under any smooth (homotopic) variation of the fields. Let us denote the fields by $\chi_{\kappa}$, and they can be scalar, vector, spinor fields, and the index $\kappa$ stands for the space-time and internal indices. The invariance of $Q$ under smooth variations of the fields lead to the identities
\br
\delta\,Q=0\quad&\rightarrow &\quad 
\frac{\delta\, {\cal A}_{\alpha}}{\delta\,\chi_{\kappa}}\, {\widetilde{\cal A}}_{\alpha}-\partial_{\mu}\(\frac{\delta\, {\cal A}_{\alpha}}{\delta\,\partial_{\mu}\chi_{\kappa}}\, {\widetilde{\cal A}}_{\alpha}\)
\lab{topidentity}\\
&+&
{\cal A}_{\alpha}\,\frac{\delta\, {\widetilde{\cal A}}_{\alpha}}{\delta\,\chi_{\kappa}} -\partial_{\mu}\({\cal A}_{\alpha}\,\frac{\delta\, {\widetilde{\cal A}}_{\alpha}}{\delta\,\partial_{\mu}\chi_{\kappa}}\)=0
\nonumber
\er
If we now impose the first order differential equations, or self-duality equations, on the fields as
\be
{\cal A}_{\alpha}=\pm {\widetilde{\cal A}}_{\alpha}
\lab{sdeqs}
\ee
it follows that together with the identities \rf{topidentity} they imply the equations 
\br
&&\frac{\delta\, {\cal A}_{\alpha}}{\delta\,\chi_{\kappa}}\, {{\cal A}}_{\alpha}-\partial_{\mu}\(\frac{\delta\, {\cal A}_{\alpha}}{\delta\,\partial_{\mu}\chi_{\kappa}}\, {{\cal A}}_{\alpha}\)
\nonumber\\
&+&
{\widetilde{\cal A}}_{\alpha}\,\frac{\delta\, {\widetilde{\cal A}}_{\alpha}}{\delta\,\chi_{\kappa}} -\partial_{\mu}\({\widetilde{\cal A}}_{\alpha}\,\frac{\delta\, {\widetilde{\cal A}}_{\alpha}}{\delta\,\partial_{\mu}\chi_{\kappa}}\)=0
\lab{eleqs}
\er
But \rf{eleqs} are the Euler-Lagrange equations associated to the functional 
\be
E=\frac{1}{2}\,\int d^dx\,\left[{\cal A}_{\alpha}^2+{\widetilde{\cal A}}_{\alpha}^2\right]
\lab{energy}
\ee
So, first order differential equations together with second order topological identities lead to second order Euler-Lagrange equations. Note that, if $E$ is positive definite then the self-dual solutions saturate a lower bound on $E$ as follows. From \rf{sdeqs} we have that ${\cal A}_{\alpha}^2={\widetilde{\cal A}}_{\alpha}^2=\pm {\cal A}_{\alpha}\,{\widetilde{\cal A}}_{\alpha}$. Therefore, if ${\cal A}_{\alpha}^2\geq 0$, and consequently ${\widetilde{\cal A}}_{\alpha}^2\geq 0$, we have that 
\br
{\cal A}_{\alpha}= {\widetilde{\cal A}}_{\alpha}\quad &\rightarrow& \quad Q=\int d^dx\, {\cal A}_{\alpha}^2\geq 0
\nonumber\\
{\cal A}_{\alpha}= -{\widetilde{\cal A}}_{\alpha}\quad &\rightarrow& \quad Q=-\int d^dx\, {\cal A}_{\alpha}^2\leq 0
\er
Therefore we have that 
\be
E=\frac{1}{2}\,\int d^dx\,\left[{\cal A}_{\alpha} \mp {\widetilde{\cal A}}_{\alpha}\right]^2 \pm \int d^dx\,{\cal A}_{\alpha} \, {\widetilde{\cal A}}_{\alpha}\geq \mid Q\mid
\lab{bound}
\ee
and the equality holds true for self-dual solutions, where we have
\be
E=\int d^dx\,{\cal A}_{\alpha}^2=\int d^dx\,{\widetilde{\cal A}}_{\alpha}^2= \mid Q\mid
\ee

The splitting  of the integrand of $Q$ as in \rf{topcharge} is quite arbitrary, but once it is chosen one can still change 
${\cal A}_{\alpha}$ and ${\widetilde{\cal A}}_{\alpha}$ by the apparently innocuous transformation 
\be
{\cal A}_{\alpha}\rightarrow {\cal A}_{\alpha}^{\prime}={\cal A}_{\beta}\,k_{\beta\,\alpha}\;;\qquad
{\widetilde{\cal A}}_{\alpha} \rightarrow {\widetilde{\cal A}}_{\alpha}^{\prime}=k^{-1}_{\alpha\,\beta}{\widetilde{\cal A}}_{\beta}
\ee
The topological charge does not change and so it is still invariant under homotopic transformations. Therefore, we can now apply the same reasoning as above with the transformed quantities ${\cal A}_{\alpha}^{\prime}$ and ${\widetilde{\cal A}}_{\alpha}^{\prime}$. The transformed self-duality equations are
\be
{\cal A}_{\beta}\,k_{\beta\,\alpha}=\pm k^{-1}_{\alpha\,\beta}{\widetilde{\cal A}}_{\beta} \quad\rightarrow \quad
{\cal A}_{\beta}\,h_{\beta\,\alpha}=\pm {\widetilde{\cal A}}_{\alpha}
\lab{newsdeqs}
\ee
where we have defined the symmetric and invertible matrix 
\be
h\equiv k\,k^T
\ee
Together with the transformed identities \rf{topidentity}, the new self-duality equations \rf{newsdeqs} imply the Euler-Lagrange equations associated to the energy
\be
E^{\prime}=\frac{1}{2}\,\int d^dx\,\left[h_{\alpha\beta}\,{\cal A}_{\alpha}\,{\cal A}_{\beta}+h^{-1}_{\alpha\beta}\,{\widetilde{\cal A}}_{\alpha}\,{\widetilde{\cal A}}_{\beta}\right]
\lab{newenergydual}
\ee
Note that the matrix $h$, or equivalently $k$, can be used to introduce new fields in the theory without changing the topological charge $Q$ and therefore its field content. In addition, the new self-duality equations \rf{newsdeqs} will also imply the Euler-Lagrange equations associated to such new fields coming from $E^{\prime}$. Indeed, if the topological charge does not depend upon these new fields, and so does not ${\cal A}_{\alpha}$ and ${\widetilde{\cal A}}_{\alpha}$, then the Euler-Lagrange equations associated to $h_{\alpha\beta}$ is ${\cal A}_{\alpha}\,{\cal A}_{\beta}-{\widetilde{\cal A}}_{\gamma}\,h^{-1}_{\gamma\alpha}\,{\widetilde{\cal A}}_{\delta}\,h^{-1}_{\delta\beta}=0$. But that follows from the self-duality equations \rf{newsdeqs}.

Note that \rf{newsdeqs} implies $h_{\alpha\beta}\,{\cal A}_{\alpha}\,{\cal A}_{\beta}=h^{-1}_{\alpha\beta}\,{\widetilde{\cal A}}_{\alpha}\,{\widetilde{\cal A}}_{\beta}=\pm {\cal A}_{\alpha}\,{\widetilde{\cal A}}_{\alpha}$. Therefore, if $h_{\alpha\beta}\,{\cal A}_{\alpha}\,{\cal A}_{\beta}\geq 0$, and consequently $h^{-1}_{\alpha\beta}\,{\widetilde{\cal A}}_{\alpha}\,{\widetilde{\cal A}}_{\beta}\geq 0$, we have that the bound follows in the same way as before 
\br
E^{\prime}&=&\frac{1}{2}\,\int d^dx\,\left[{\cal A}_{\beta}\,k_{\beta\,\alpha}\mp k^{-1}_{\alpha\,\beta}{\widetilde{\cal A}}_{\beta} \right]^2\pm \int d^dx\,{\cal A}_{\alpha}\,{\widetilde{\cal A}}_{\alpha} 
\nonumber\\
&\geq& \mid Q\mid
\er
Such ideas have been applied quite successfully in many Skyrme type models \cite{bpswojtek,bpsshnir,laf2017,us} and in two dimensional scalar field theories \cite{sd2d}.

\section{Self-Duality in the Yang-Mills-Higgs System}
\label{sec:ymh}
\setcounter{equation}{0}

We now consider a Yang-Mills theory for a gauge group $G$ coupled to a Higgs field in the adjoint representation of $G$. The relevant topological charge is the magnetic charge
\be
Q_M=\int_{\IR^3} d^3x\,\partial_i \trace\( B_i\, \Phi\)=\int_{S^2_{\infty}} d\Sigma_i\, \trace\( B_i\, \Phi\)
\lab{magtopcharge}
\ee
where
\br
B_i&=&-\frac{1}{2}\,\ve_{ijk}\,F_{jk}
\lab{bfdef}\\
 F_{ij}&=&\partial_iA_j-\partial_jA_i+i\,e\,\sbr{A_i}{A_j}
\nonumber
\er
and $A_i=A_i^a\,T_a$, $\Phi=\Phi_a\,T_a$, with $T_a$, $a=1,2,\ldots {\rm dim}\,G$, being a basis of the Lie algebra of the gauge group $G$, satisfying $\sbr{T_a}{T_b}=i\,f_{abc}\,T_c$, and ${\rm Tr}\(T_a\,T_b\)=\kappa\, \delta_{ab}$, and $\kappa$ being the Dynkin index of the representation where the trace is taken. In \rf{magtopcharge} we have used the normalised trace $\trace\equiv \frac{1}{\kappa}\,{\rm Tr}$.  
Adding to the integrand in \rf{magtopcharge} the trivially vanishing term $\trace\(\sbr{A_{i}}{B_i\,\Phi}\)$, and using the Bianchi identity $D_i\,B_i=0$, with $D_i *=\partial_i*+i\,e\,\sbr{A_i}{*}$, one can write \rf{magtopcharge} as
\be
Q_M=\int_{\IR^3} d^3x\,\trace\(B_i\,D_i\Phi\)=\int_{\IR^3} d^3x\,B_i^a\,\(D_i\Phi\)^a
\lab{topchargeymh}
\ee
Following the ideas described in section \ref{sec:sd}, we shall split the integrand of such a topological charge as \cite{foot} 
\be
{\cal A}_{\alpha}\equiv B_i^b\,k_{ba}\;;\qquad {\widetilde{{\cal A}}_{\alpha}}\equiv k^{-1}_{ab}\,\(D_i\Phi\)^b
\ee
and the self-duality equations are then given by
\be
B_i^b\,h_{ba}=\eta\,\(D_i\Phi\)^a\;;\qquad \eta=\pm 1\;;\qquad h=k\,k^T
\lab{sdeqsymh}
\ee
The static energy of our generalised Yang-Mills-Higgs system, according to \rf{newenergydual},  is given by
\be
E_{YMH}=\frac{1}{2}\,\int d^3x\,\left[h_{ab}\,B_i^a\,B_i^b+h^{-1}_{ab}\,\(D_i\Phi\)^a\,\(D_i\Phi\)^b\right]
\lab{energyymh}
\ee
For the solutions of the self-duality equations we have that
\be
E_{YMH}= Q_M
\lab{energytopchargerel}
\ee
The four dimensional action associated to \rf{energyymh} is 
\br
S_{YMH}&=&\int d^4x\,\left[-\frac{1}{4}\,h_{ab}\,F_{\mu\nu}^a\,F^{b\,\mu\nu}
\right. \nonumber\\
&+&\left.\frac{1}{2}\,h^{-1}_{ab}\,\(D_{\mu}\Phi\)^a\,\(D^{\mu}\Phi\)^b\right]
\lab{actionymh}
\er
Under a gauge transformation $A_{\mu}\rightarrow g\,A_{\mu}\,g^{-1}+\frac{i}{e}\,\partial_{\mu}g\,g^{-1}$, we have that
$F_{\mu\nu}\rightarrow g\,F_{\mu\nu}\,g^{-1}$ and $D_{\mu}\Phi\rightarrow g\,D_{\mu}\Phi\,g^{-1}$. Therefore, the action \rf{actionymh}, the energy \rf{energyymh} and the self-duality equations \rf{sdeqsymh} are invariant under
\br
F_{\mu\nu}^a&\rightarrow& d_{ab}\(g\)\,F_{\mu\nu}^b\,;\qquad 
\(D_{\mu}\Phi\)^a\rightarrow d_{ab}\(g\)\,\(D_{\mu}\Phi\)^b
\nonumber\\
h_{ab}&\rightarrow& d_{ac}\(g\)\,d_{bd}\(g\)\,h_{cd}
\lab{gaugetransformgen}
\er
where $d\(g\)$ are the matrices of the adjoint representation of the gauge group
\be
g\,T_a\,g^{-1}=T_b\,d_{ba}\(g\)
\ee
The adjoint representation of a compact simple Lie group is unitary and real, and so its matrices are orthogonal, i.e. $d\,d^T=\one$. The action \rf{actionymh} is Lorentz invariant in the four dimensional Minkowski space-time. However, the static energy \rf{energyymh} and the self-duality equations \rf{sdeqsymh} are conformally invariant in the three dimensional space, as we show in the appendix \ref{app:conformal}. 

Note that under space parity $x_i\rightarrow -x_i$, and $t\rightarrow t$, we have that $A_i\rightarrow - A_i$, and $A_0\rightarrow A_0$, and so $B_i\rightarrow B_i$. Therefore, the self-duality equations \rf{sdeqsymh} are invariant under space parity if the Higgs fields $\Phi^a$ are pseudo-scalars and the fields $h_{ab}$ are scalars, and consequently the energy \rf{energyymh} and the topological charge \rf{topchargeymh} are parity invariant. However, if the Higgs fields are scalars and $h_{ab}$ are pseudo-scalars, the self-duality equations are still invariant but both, the energy and topological charge, change sign under parity.  Perhaps, the most sensible situation to assume is that one where both the Higgs and $h$ fields are scalars, and so the self-duality equations are not invariant. In that case, the energy \rf{energyymh} is parity invariant, but the  topological charge \rf{topchargeymh} changes sign.  Therefore, space parity would map self-dual solutions into anti-self-dual solutions.

The fields of our model are the gauge fields $A_{\mu}^a$, the Higgs fields $\Phi^a$, and the scalar fields $h_{ab}$. The static Euler-Lagrange equations associated to those fields, following from \rf{actionymh}, or equivalently \rf{energyymh}, are
\br
D_i\(h\,F_{ij}\)&=&i\,e\,\sbr{\Phi}{h^{-1}\,D_j\Phi}
\lab{eqa}
\\
D_i\(h^{-1}\,D_i\Phi\)&=&0
\lab{eqphi}
\\
B_i^a\,B_i^b&=&h^{-1}_{ac}\,h^{-1}_{bd}\,\(D_i\Phi\)^c\,\(D_i\Phi\)^b
\lab{eqh}
\er
where we have introduced the notation
\be
h\,F_{ij}\equiv T_a\,h_{ab}\,F_{ij}^b\,;\qquad h^{-1}\,D_i\Phi\equiv T_a\,h^{-1}_{ab}\,\(D_i\Phi\)^b
\lab{hftdef}
\ee
Note that we can write \rf{sdeqsymh} as
\be
B_i^a=\eta\, \(D_i\Phi\)^c\,h^{-1}_{ca}
\lab{sdeqsymh2}
\ee
and contracting with $B_i^b$ we get
\be
\tau_{ab}=\eta\, \sigma_{ac}\,h^{-1}_{cb}
\ee
with
\be
\tau_{ab}\equiv B_i^a\,B_i^b\,;\qquad
\sigma_{ab}\equiv B_i^a\,\(D_i\Phi\)^b
\lab{tausigmadef}
\ee
and these matrices will be important in what follows. We can now write \rf{eqh} as
\br
&&B_i^a\,B_i^b-h^{-1}_{ac}\,h^{-1}_{bd}\,\(D_i\Phi\)^c\,\(D_i\Phi\)^d=
\nonumber\\
&&\left[B_i^a-h^{-1}_{ac}\,\(D_i\Phi\)^c\right]\left[B_i^b+h^{-1}_{bd}\,\(D_i\Phi\)^d\right]
\nonumber\\
&+&\(\sigma\,h^{-1}\)_{ba}-\(\sigma\,h^{-1}\)_{ab}
\lab{checkeqh}
\er
Therefore, using \rf{sdeqsymh2} and \rf{tausigmadef} one observes that the r.h.s. of \rf{checkeqh} vanishes, and so the self-duality equations \rf{sdeqsymh} do imply the Euler-Lagrange equations \rf{eqh} for the $h$ fields.  Contracting both sides of \rf{sdeqsymh2} with $T_a$, and taking the covariant divergency of its both sides one gets, using \rf{bfdef} and \rf{hftdef}, 
\be
-\frac{1}{2}\,\ve_{ijk}\,D_i\,F_{jk}=\eta\, D_i\(h^{-1}\,D_i\Phi\)
\lab{checkeqphi}
\ee
But the l.h.s of \rf{checkeqphi} is the Bianchi identity and so it vanishes. Therefore, the self-duality equations \rf{sdeqsymh2} imply the Euler-Lagrange equations \rf{eqphi} for the Higgs field $\Phi$.

Using the notation of \rf{hftdef} and \rf{bfdef} we can write \rf{sdeqsymh} as $h\,F_{ij}=-\eta\, \ve_{ijk}\,D_k\Phi$. Taking the covariant divergence on both sides one gets $D_i\(h\,F_{ij}\)=-\eta\, i\,e\,\sbr{B_j}{\Phi}$, where we have used the Jacobi identity. Contracting \rf{sdeqsymh2} with $T_a$, commuting both sides with $\Phi$, and using the notation of \rf{hftdef}, we get $\sbr{\Phi}{B_j}=\eta\, \sbr{\Phi}{h^{-1}\,D_j\Phi}$. Therefore, we observe that the self-duality equations imply the Euler-Lagrange equations \rf{eqa} for the gauge fields $A_i$. So, the solutions of the self-duality equations also solve all three Euler-Lagrange equations \rf{eqa}, \rf{eqphi} and \rf{eqh}. 

Since the matrix $h$ is always invertible, we note from \rf{tausigmadef} that the matrix $\tau$ is invertible whenever $\sigma$ is invertible and vice-versa. Therefore, on the regions of $\IR^3$ where the matrix $\tau$ is invertible, we can use the self-duality equations, or equivalently \rf{tausigmadef}, to write the matrix of the $h$-fields in terms of the gauge and Higgs fields as
\be
h=\eta\,\tau^{-1}\,\sigma
\lab{hsol}
\ee 
Such a relation means that whenever $\tau$ is invertible the self-duality equations are automatically satisfied by an $h$ matrix given by \rf{hsol}, and so the $h$-fields are just spectators in the sense that they adjust themselves to the given $\Phi$ and $A_i$ field configurations to solve the self-duality equations. 

Note in addition that, since $\tau$ and $h$ are symmetric, it follows that $\tau\,h=\eta\,\sigma$ and $h\,\tau=\eta\,\sigma^T$. Therefore, $\sbr{\tau}{h}=\eta\(\sigma-\sigma^T\)$. So, $\sigma$ will be symmetric whenever $\tau$ and $h$ commute. 

\subsection{The $h$-fields}

Note from \rf{gaugetransformgen} that the $h$-fields transform under gauge transformations as $h\rightarrow d\(g\)\,h\,d^T\(g\)$, with $d\,d^T=\one$, and so the eigenvalues of $h$ are gauge invariant. Since $h$ is a symmetric and real matrix, it can be diagonalized by an orthogonal transformation 
\be
h=M\, h_D\,M^T\;;\qquad M\,M^T=\one\;;\qquad \(h_D\)_{ab}=\lambda_a\,\delta_{ab}
\lab{diaghgeneral}
\ee
Therefore, it is convenient to split the ${\cal N}\({\cal N}+1\)/2$ $h$-fields, where ${\cal N}$ is the dimension of the gauge group $G$, into two sets. The first set contains the ${\cal N}$ gauge in\-va\-riant $\lambda$-fields, and the second set contains the ${\cal N}\({\cal N}-1\)/2$ fields parameterizing the orthogonal matrix $M$. Accor\-ding to \rf{gaugetransformgen}, under a gauge transformation, such fields transform as
\be
\lambda_a \rightarrow \lambda_a\;;\qquad\qquad M\rightarrow d\(g\)\, M
\lab{gaugetransformlbdm}
\ee 
Under a conformal transformation in the three dimensional space $\IR^3$, as described in Appendix \ref{app:conformal}, we have that such fields transform as
\be
\delta \lambda_a=\Omega\, \lambda_a\;;\qquad\qquad\qquad \delta M=0
\ee
We now introduce the quantites
\be
{\cal F}_{\mu\nu}^a\equiv M_{ab}^T\,F_{\mu\nu}^b\;;\qquad
\({\cal D}_{\mu}\Phi\)_a\equiv M_{ab}^T\,\(D_{\mu}\Phi\)_b
\lab{fcaldef}
\ee
From \rf{gaugetransformgen} and \rf{gaugetransformlbdm} one observes that such quantities are gauge invariant,  i.e. 
\be
{\cal F}_{\mu\nu}^a\rightarrow {\cal F}_{\mu\nu}^b\;;\qquad 
\({\cal D}_{\mu}\Phi\)_a \rightarrow \({\cal D}_{\mu}\Phi\)_b
\ee
Therefore, the four dimensional action \rf{actionymh} and static energy \rf{energyymh} can be written solely in terms of  gauge invariant quantities as
\br
S_{YMH}&=&\int d^4x\,\left[-\frac{1}{4}\,\lambda_{a}\,{\cal F}_{\mu\nu}^a\,{\cal F}^{a\,\mu\nu}
\right. \nonumber\\
&+&\left.\frac{1}{2}\,\frac{1}{\lambda_a}\,\({\cal D}_{\mu}\Phi\)^a\,\({\cal D}^{\mu}\Phi\)^a\right]
\lab{actionymhinv}
\er
and 
\be
E_{YMH}=\frac{1}{2}\,\int d^3x\,\left[\lambda_{a}\,{\cal B}_i^a\,{\cal B}_i^a+\frac{1}{\lambda_a}\,\({\cal D}_i\Phi\)^a\,\({\cal D}_i\Phi\)^a\right]
\lab{energyymhinv}
\ee
where, following \rf{bfdef}, we have denoted 
\be
{\cal B}_i^a=-\frac{1}{2}\,\ve_{ijk}\,{\cal F}_{jk}^a
\lab{bfdefinv}
\ee
The self-duality equations \rf{bfdef}, can also be written in terms of  gauge invariant quantities only
\be
{\cal B}_i^a\,\lambda_{a}=\eta\,\({\cal D}_i\Phi\)^a\;;\qquad \qquad \eta=\pm 1
\lab{sdeqsymhinv}
\ee

It is interesting to note that there is a standard way of constructing quantities, out of the field tensor and the covariant derivative of an adjoint Higgs field, that transform {\em globally} under {\em local}  gauge transformations, using the Wilson line. Given a curve $x^{\mu}\(\sigma\)$ on space-time, parameterized by $\sigma$, the Wilson line operator $W$ is defined through the differential equation 
\be
\frac{d\,W}{d\,\sigma}+i\,e\,A_{\mu}\,\frac{d\,x^{\mu}}{d\,\sigma}\,W=0
\lab{weqdef}
\ee
Under a gauge transformation $A_{\mu}\rightarrow g\,A_{\mu}\,g^{-1}+\frac{i}{e}\,\partial_{\mu}g\,g^{-1}$, the Wilson line transforms as
\be
W\rightarrow g_f\, W\,g_i^{-1}
\ee
where $g_f$ and $g_i$ are the group elements at the final and initial points respectively of the curve $x^{\mu}\(\sigma\)$. Consider now the quantities
\be
F^W_{\mu\nu}\equiv W^{-1}\,F_{\mu\nu}\,W\;;\qquad \(D_{\mu}\Phi\)^W\equiv W^{-1}\,D_{\mu}\Phi\, W
\lab{conjugateddef}
\ee
where the Wilson line is defined on a curve that ends at the point where $F_{\mu\nu}$ and $D_{\mu}\Phi$ are evaluated. Therefore, under a gauge transformation, such quantities transform as
\be
F^W_{\mu\nu}\rightarrow g_i\, F^W_{\mu\nu}\, g_i^{-1}\;;\qquad 
\(D_{\mu}\Phi\)^W\rightarrow g_i\,\(D_{\mu}\Phi\)^W\,g_i^{-1}
\lab{gaugetransformfw}
\ee
If we now restrict ourselves to the case where all curves start at given fixed reference point, it turns out that $g_i$ is a fixed element of $G$, and so the conjugated quantities $F_{\mu\nu}^W$ and $\(D_{\mu}\Phi\)^W$ transform under global gauge transformations. Note that \rf{weqdef} is a first order differential equation and so the Wilson line is defined up to an integration constant, i.e. if $W$ is a solution of \rf{weqdef}, so it is $W\,W_0$, with $W_0$ a constant group element. Note that $W_0$ is the value of the Wilson line at the initial point of the curve. Therefore, the global gauge transformations of the quantities $F_{\mu\nu}^W$ and $\(D_{\mu}\Phi\)^W$ amounts to the freedom of the choice of such integration constant. 

The field tensor conjugated by the Wilson line appears in the usual non-abelian Stokes theorem, as well as in its generalizations to two-form connections \cite{afg1,afg2}. Such theorems were used to construct the integral form of the Yang-Mills equations in \cite{ym1,ym2}. These integral  equations are expressed in terms of the field tensor and its Hodge dual, conjugated by the Wilson line in the way explained above. In addition, the Wilson lines have to be evaluated on curves all starting at a fixed reference point, and  the integration constants associated to the Wilson line have to be restricted to the center of the gauge group in order for the integral equations to be gauge covariant \cite{ym1,ym2}. So, in order to keep the integration constant in the center of the group, we have to take $g_i$ in \rf{gaugetransformfw} also in the center, and then $F^W_{\mu\nu}$ and $\(D_{\mu}\Phi\)^W$ are gauge invariant like ${\cal F}_{\mu\nu}^a$ and $\({\cal D}_{\mu}\Phi\)_a$, given in \rf{fcaldef}. 

From \rf{conjugateddef} we have that
\br
F^W_{\mu\nu}&=& F_{\mu\nu}^a\,W^{-1}\,T_a\,W= F_{\mu\nu}^a\,T_b\,d_{ba}\(W^{-1}\)
\nonumber\\
&=&T_b\,d^T_{ba}\(W\)\,F_{\mu\nu}^a
\er
and similarly for $\(D_{\mu}\Phi\)^W$. Therefore, we have that
\br
\(F^W_{\mu\nu}\)^a&=&d^T_{ab}\(W\)\,F_{\mu\nu}^b
 \lab{conjugateddef2}\\
\left[ \(D_{\mu}\Phi\)^W\right]^a&=&d^T_{ab}\(W\)\,\(D_{\mu}\Phi\)^b
\nonumber
\er

The covariant derivative of the $M$-fields is $D_{\mu}M=\partial_{\mu}M+i\,e\,d\(A_{\mu}\)\,M$, since it transforms as $D_{\mu}M\rightarrow d\(g\)\,D_{\mu}M$, and so in the same way as $M$ in \rf{gaugetransformlbdm}. Given  a curve $x^{\mu}\(\sigma\)$, consider the quantity
\be
\frac{d\,x^{\mu}}{d\,\sigma}\,D_{\mu}M=\frac{d\,M}{d\,\sigma}+i\,e\,d\(A_{\mu}\)\,\frac{d\,x^{\mu}}{d\,\sigma}\,M
\lab{covariantdermcurve}
\ee
One observes that, in the regions of space where $D_{\mu}M=0$ or where $D_{\mu}M$ is perpendicular to the curve, the matrix $M$ satisfies the same equation as the Wilson line $W$ in the adjoint representation, i.e. \rf{weqdef}. Therefore for curves on those regions we have that $M=d\(W\)$, and so the quantities $\({\cal F}_{\mu\nu}^a\,,\, \({\cal D}_{\mu}\Phi\)_a\)$,  and $\(\(F^W_{\mu\nu}\)^a\,,\, \left[ \(D_{\mu}\Phi\)^W\right]^a\)$, given in \rf{fcaldef} and \rf{conjugateddef2} respectively, are the same. In the examples that we discuss below, we show that, for the cases where the matrix $h$ is completely determined in terms of the gauge fields, as given in \rf{hsol}, it is possible to choose curves  $x^{\mu}\(\sigma\)$, starting at a fixed reference point and ending at any point of $\IR^3$, such that  $M=d\(W\)$. 

The $h$-fields are constituted of two distinct type of fields. The $\lambda$-fields, according to \rf {gaugetransformlbdm}and \rf{diaghgeneral}, are gauge invariant and have conformal weight one. Therefore, they are like dilaton fields and are responsible for the conformal invariance of the theory \rf{actionymh} in the three dimensional space $\IR^3$. Dilaton fields have been introduce in effective theories for Yang-Mills theories in relation to the trace anomaly \cite{migdal,ellis}. The dilaton field is related to the expectation value of the trace of the energy-momentum tensor, or equivalently to the gluon condensate, and it couples to the Yang-Mills Lagrangian in a way similar to the coupling of the $\lambda$-fields in \rf{energyymhinv}. In such a context our theory \rf{actionymh} can be seen as an effective field theory. The $M$-fields on the other hand are scalars under the conformal group and transform under gauge transformations in a way similar to the Wilson line operator in the adjoint representation. As they lead to dressed quantities like \rf{fcaldef}, which are gauge invariant, and also given their close relation to the   Wilson operator, which plays a role in the low energy regime of Yang-Mills, they reinforce the interpretation of the theory \rf{actionymh} as an effective Yang-Mills-Higgs  theory.

\section{Spherically Symmetric Solutions for  $G=SU(2)$}
\label{sec:su2}
\setcounter{equation}{0}

We use the spherical ans\"atz of 't Hooft-Polyakov given by \cite{thooft,polyakov}
\br
\Phi&=& \frac{1}{e}\,\frac{H\(r\)}{r}\,{\hat r}_a T_a
\nonumber\\
A_i&=& -\frac{1}{e}\,\ve_{ija}\, \frac{{\hat r}_j}{r}\,\(1-K\(r\)\)\,T_a
\lab{thooftansatz}\\
A_0&=&0
\nonumber
\er
with ${\hat r}_i=x_i/r$, and $T_a$, $a=1,2,3$, being the basis of the $SU(2)$ Lie algebra satisfying $\sbr{T_a}{T_b}=i\,\ve_{abc}\,T_c$. We then get that
\br
B_i&=& B_i^a\,T_a\,;\quad B_i^a=\frac{1}{e\,r^2}\left[ r\,K^{\prime}\,\Omega_{ia}+ \(K^2-1\)\,\Lambda_{ia}\right]
\nonumber\\
D_i\Phi&=&\(D_i\Phi\)^a\,T_a
\lab{bdphilambda}\\
 \(D_i\Phi\)^a&=&\frac{1}{e\,r^2}\left[H\,K\,\Omega_{ia}+\(r\, H^{\prime}-H\)\,\Lambda_{ia}\right]
\nonumber
\er
where we have defined $\Omega\equiv \one - \Lambda$, with $\Lambda_{ab}\equiv {\hat r}_a\,{\hat r}_b$, and so $\Lambda^2=\Lambda$, $\Omega^2=\Omega$, and $\Lambda\,\Omega=\Omega\,\Lambda=0$. Therefore, the matrix $h$ that solves the self-duality equations \rf{sdeqsymh} is given by 
\be
h=\eta\left[\frac{K\,H}{r\,K^{\prime}}\, \Omega+\frac{r\,H^{\prime}-H}{\(K^2-1\)}\,\Lambda\right]
\lab{hlambda}
\ee
Note that, given any field configuration for the gauge and Higgs fields, in the ans\"atz \rf{thooftansatz}, we solve the self-duality equations with the matrix $h$ given in \rf{hlambda}, for any profile functions $H$ and $K$, as long as the eigenvalues of $h$ do not vanish. So the $h$-fields act like spectators adjusting themselves to the gauge and Higgs fields configurations.  

From \rf{tausigmadef} and \rf{bdphilambda} we then get 
\be
\tau= \frac{1}{e^2\,r^4}\left[\(r\,K^{\prime}\)^2\,\Omega+\(K^2-1\)^2\,\Lambda\right]
\lab{taulambda}
\ee
and
\be
\sigma= \frac{1}{e^2\,r^4}\left[r\,K^{\prime}\,K\,H\,\Omega+\(K^2-1\)\(r\,H^{\prime}-H\)\,\Lambda\right]
\lab{sigmalambda}
\ee
 Therefore, the matrix $\sigma$ is also symmetric. In addition, any two matrices that are linear combinations of $\Lambda$ and $\Omega$, commute among themselves. So, $\sbr{\tau}{\sigma}=0$. Note that, for any matrix of the form $L=\alpha\, \Omega+\beta\,\Lambda$, its inverse is simply $L^{-1}= \Omega/\alpha+\Lambda/\beta$. 
 
Note that $\Lambda$ has a zero eigenvalue twice degenerated, and a single eigenvalue unity. The eingenvector corresponding to the unity eigenvalue is clearly
\br
v_a^{(3)}={\hat r}_a\;;\qquad {\rm or}\qquad 
v^{(3)}=\(\begin{array}{c}
\sin\theta\,\cos\phi\\
\sin\theta\,\sin\phi\\
\cos\theta
\end{array}\)
\lab{v3def}
\er
where $\theta$ and $\phi$ are the angles of the spherical polar coordinates. 
We take the basis for the degenerated zero eigenvalue subspace as
\br
v^{(1)}=\(\begin{array}{c}
\cos\theta\,\cos\phi\\
\cos\theta\,\sin\phi\\
-\sin\theta
\end{array}\);\quad
v^{(2)}=\(\begin{array}{c}
-\sin\phi\\
\cos\phi\\
0
\end{array}\)
\lab{zereigenvectors}
\er
and so
\be
\Lambda\cdot v^{(1)}=\Lambda\cdot v^{(2)}=0;\quad \Lambda\cdot v^{(3)}=v^{(3)};\quad v^{(a)}\cdot v^{(b)}=\delta_{ab}
\ee
Clearly, those three vectors are eigenvectors of $\Omega$ with eigenvalues $1$ (doubly degenerate) and zero respectively.  Therefore, for a matrix of the form $L=\alpha\, \Omega+\beta\,\Lambda$, the eigenvalues are $\(\alpha\,,\,\alpha\,,\,\beta\)$, and so the eigenvalues  of $h$, $\tau$ and $\sigma$, can be read off directly from their expressions \rf{hlambda}, \rf{taulambda} and \rf{sigmalambda}. Those matrices can be simultaneously diagonalised by an orthogonal matrix $M$, i.e.
\br
h&=&M\,h_D\, M^T\,;\qquad\quad \tau=M\,\tau_D\, M^T
\nonumber\\
 \sigma&=&M\,\sigma_D\, M^T\,;  \qquad\quad M\,M^T=\one
 \lab{htausigmadiagonal}
\er
with
\br
h_D&=&{\rm diag.}\(\lambda_1\,,\,\lambda_1\,,\,\lambda_2\)
\nonumber\\
\tau_D&=&{\rm diag.}\(\omega_1\,,\,\omega_1\,,\,\omega_2\)
\lab{htausigmadiag}\\
\sigma_D&=&{\rm diag.}\(\eta\,\lambda_1\,\omega_1\,,\,\eta\,\lambda_1\,\omega_1\,,\,\eta\,\lambda_2\,\omega_2\)
\nonumber
\er
with
\br
\lambda_1&=&\eta\,\frac{K\,H}{r\,K^{\prime}}\,;\qquad\lambda_2=\eta\,\frac{\(r\,H^{\prime}-H\)}{\(K^2-1\)}
\lab{htausigmadiag2}\\
\omega_1&=&\frac{1}{e^2\,r^4}\,\(r\,K^{\prime}\)^2\,;\qquad 
\omega_2=\frac{1}{e^2\,r^4}\,\(K^2-1\)^2
\nonumber
\er

\subsection{The usual BPS monopole}

Note that the matrix $h$, given in \rf{hlambda}, will be the unity matrix  whenever the coefficients of $\Omega$ and $\Lambda$ are both equal to the sign $\eta=\pm 1$, i.e.
\be
h=\one\quad\rightarrow\quad r\,K^{\prime}=\eta\, K\,H\,;\qquad r\,H^{\prime}-H=\eta\,\(K^2-1\)
\lab{husualbps}
 \ee
and those are the self-duality equations for the profile functions of the 't Hooft-Polyakov ans\"atz for the Bogomolny-Prasad-Sommerfield (BPS) monopole \cite{bogo,prasad}. The solution is given by
\be
H=-\eta\,\left[\xi\,\coth\(\xi\)-1\right]\,;\qquad K=-\eta\,\frac{\xi}{\sinh\(\xi\)}
\lab{ordinarybpssol}
\ee
with $ \xi=r/r_0$, and $r_0$ being an arbitrary length scale. 

\subsection{The 't Hooft-Polyakov monopole}

In the case of the  't Hooft-Polyakov monopole \cite{thooft,polyakov}, the profile functions of the ans\"atz \rf{thooftansatz} satisfy
\br
\xi^2\,K^{\prime\prime}&=&K\,H^2+K\(K^2-1\)
\nonumber\\
\xi^2\,H^{\prime\prime}&=&2\,K^2\,H+\frac{\kappa}{e^2}\,H\,\(H^2-\xi^2\)
\lab{thoofteqs}
\er
where again $\xi=r/r_0$, and $\kappa$ is the parameter of the Higgs potential $V=\frac{\kappa}{4}\({\rm Tr}\Phi^2-\langle\Phi\rangle^2\)^2$, with $\langle\Phi\rangle$ being the vacuum expectation value of the Higgs field. 

The asymptotic behavior of the profile functions at infinity and at the origin are given by
\be
K\sim e^{-\xi}\;;\quad H-\xi\sim e^{-\frac{\sqrt{2\,\kappa}}{e}\, \xi}\;;\quad {\rm for}\quad \xi\rightarrow \infty
\ee
and
\be
K\sim 1\;;\qquad\quad \frac{H}{\xi}\sim 0\;; \qquad\quad {\rm for}\quad \xi\rightarrow 0
\ee
Therefore, the eigenvalues of $h$, given in \rf{htausigmadiag}, behave as
\be
\lambda_1\rightarrow -\eta\;;\qquad \lambda_2\rightarrow 0\;;\qquad {\rm for}\quad \xi\rightarrow \infty
\ee
and
\be
\lambda_1\rightarrow -\eta\,\beta\;;\qquad \lambda_2\rightarrow -\eta\,\beta\;;\qquad {\rm for}\quad \xi\rightarrow 0
\ee
with $\beta$ being a positive constant depending upon $\kappa/e^2$. Therefore, the 't Hooft-Polyakov monopole must belong to the self-dual sector corresponding to $\eta=-1$, in order to have the eigenvalues of $h$ positive, and so the static energy \rf{energyymh} positive. 

\begin{figure*}
\centering
		\includegraphics[scale=0.5]{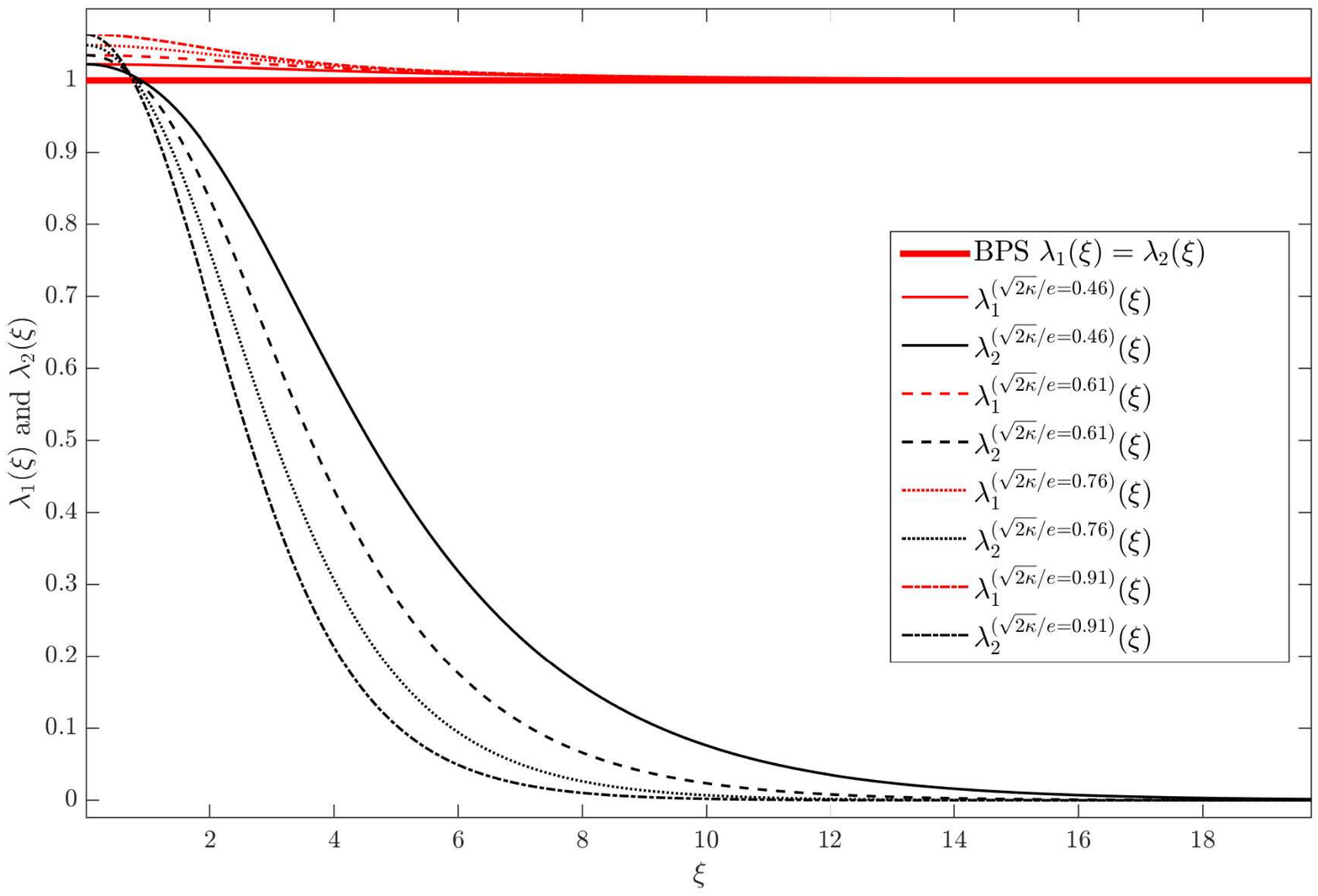}
		\caption{The eigenvalues $\lambda_1$ and $\lambda_2$, given in \rf{htausigmadiag}, for the solutions of \rf{thoofteqs} of the 't Hooft-Polyakov monopole, for some values of the parameter $\kappa/e^2$.}
		\label{fig:thooft}
\centering 
\end{figure*}

We plot in Figure \ref{fig:thooft} the eigenvalues of $h$, against $\xi$, for the 't Hooft-Polyakov monopole, for some values of  $\kappa/e^2$. Note that, at spatial infinity the eigenvalue $\lambda_1$ tend to unity, i.e. the value it has in the usual self-dual solution, given in \rf{husualbps} and\rf{ordinarybpssol}, but $\lambda_2$ tend to zero instead. It is such a different behavior of the scalar fields $h_{ab}$ that allows the configuration of the 't Hooft-Polyakov monopole to be a self-dual solution in such modified Yang-Mills-Higgs theory. 

In fact, if we turn the arguments around, we could interpret the $h$-fields as introducing a dielectric medium in the Yang-Mills-Higgs system, on the lines of what has been attempted in \cite{dahora,dahora2,bazeia1,bazeia2,zhang}. Therefore, the coupling of such a medium to the gauge and Higgs fields  replaces the Higgs potential and sustain the 'tHooft-Polyakov monopole as a solution of a self-dual theory. Instead of introducing such an structure as an external and rigid  medium, we do it dynamically through the coupling of the (non-propagating) $h$-fields to the gauge and Higgs fields.  

\subsection{Some special choices of monopole solutions}

As we have seen, any choice of profile functions $H$ and $K$, satisfying appropriate boundary conditions, leads to monopole solutions with non-trivial topological charges. We present here some monopole solutions where the eigenvalues of $h$ behave, close to the origin, in the same way as the ordinary BPS solution \rf{ordinarybpssol}, i.e.  
\be
\lambda_a \rightarrow 1\; ; \qquad  a=1,2\;;\qquad{\rm for}\quad \xi\rightarrow 0
\ee
and at infinity such eigenvalues behave in the same way as the 't Hooft-Polyakov monopole solution, i.e
\be
\lambda_1\rightarrow 1\;;\qquad \lambda_2\rightarrow 0\;;\qquad {\rm for}\quad r\rightarrow \infty
\ee
In order to do that we take the following ans\"atz for the eigenvalues $\lambda_a$ 
\be
\lambda_1= 1+ \frac{H\,K}{\xi}\;;\qquad \lambda_2=1-\(\frac{H}{\xi}\)^{\alpha}
\lab{newansatz}
\ee
with $\alpha$ a constant parameter. The ans\"atz \rf{newansatz} constitutes in fact a generalization of the one used in \cite{dahora}. Therefore, from \rf{htausigmadiag} we get the following first order differential equations for the profile functions
\br
 K^{\prime}&=&\eta\, \frac{K\,H/\xi}{\(1+K\,H/\xi\)}
 \lab{thridansatzeq} \\
  \(\frac{H}{\xi}\)^{\prime}&=&\frac{\eta}{\xi^2}\,\(K^2-1\)\,\(1-\(\frac{H}{\xi}\)^{\alpha}\)
\nonumber
\er

\begin{figure*}
\centering
		\includegraphics[scale=0.5]{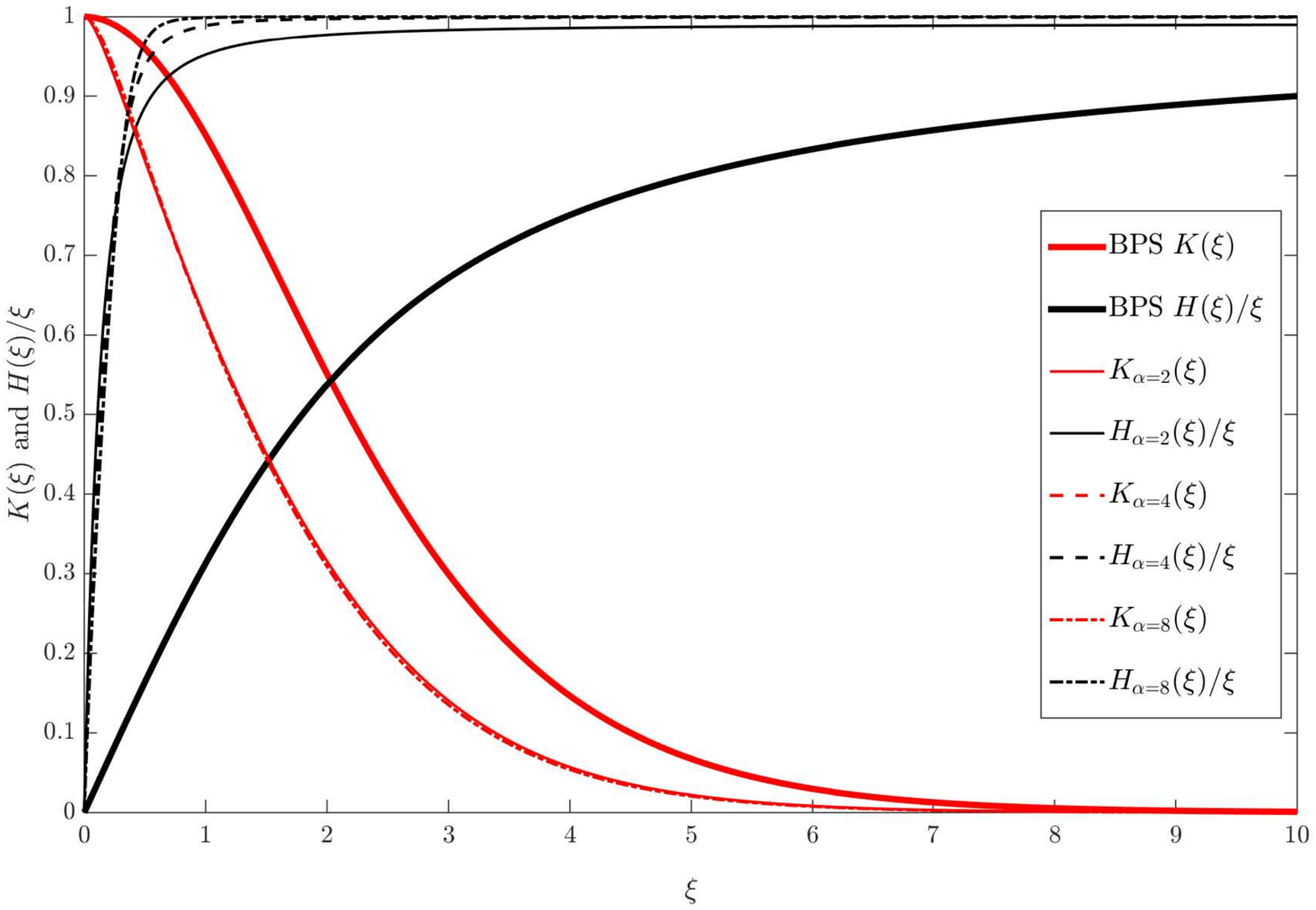}
		\caption{The profile functions $K$ and $H/\xi$, solving equations \rf{thridansatzeq}, for some values of $\alpha$, and the same functions for the usual BPS case, given in \rf{ordinarybpssol}. }
		\label{fig:thirdansatz}
\centering 
\end{figure*}

\begin{figure*}
\centering
		\includegraphics[scale=0.5]{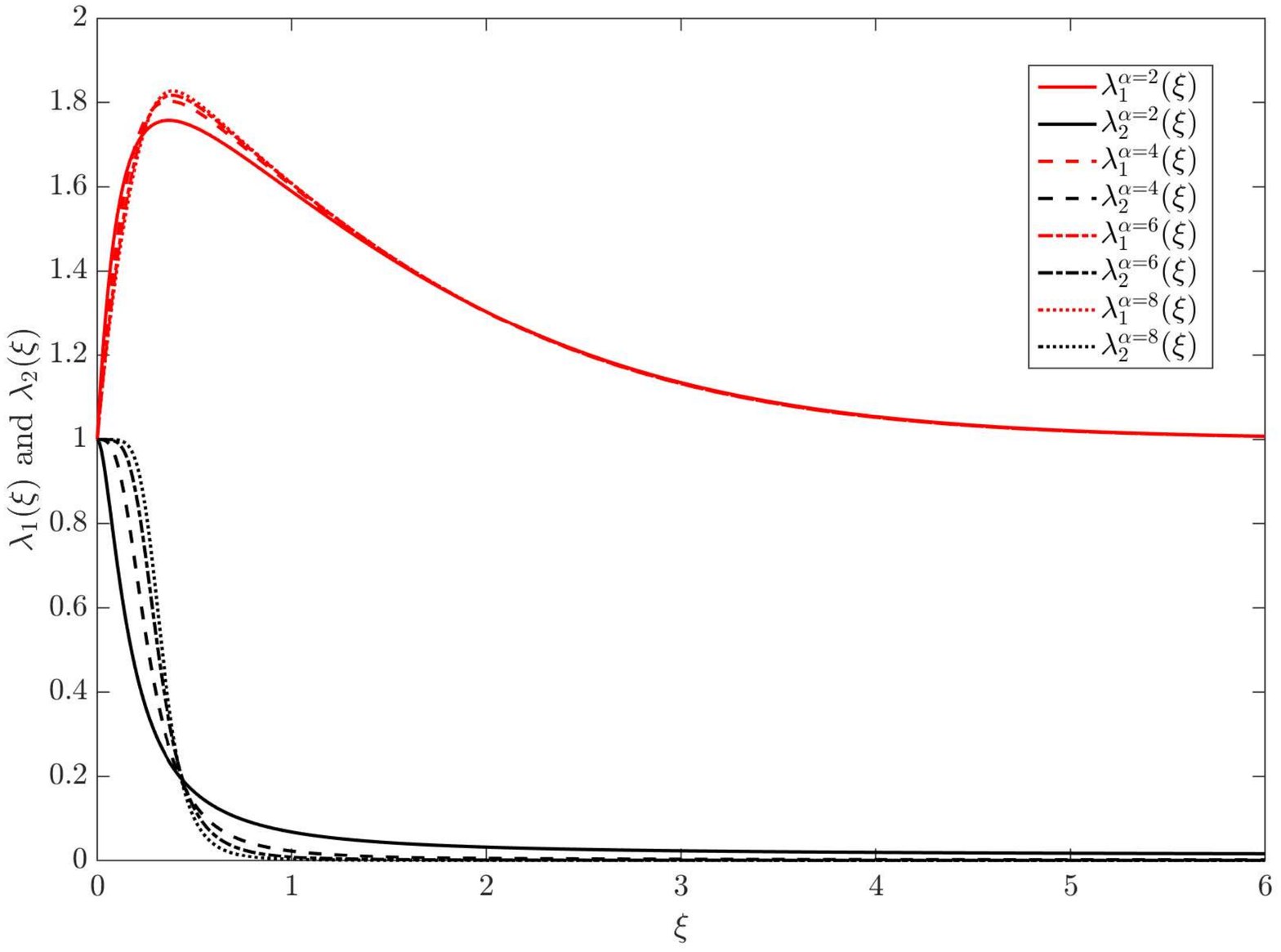}
		\caption{The eigenvalues $\lambda_a$, $a=1,2$, defined in \rf{htausigmadiag2}, for solutions of the  equations \rf{thridansatzeq}, for some values of $\alpha$.  }
		\label{fig:thirdansatz2}
\centering 
\end{figure*}

We plot in Figure \ref{fig:thirdansatz} the profile functions $K$ and $H/\xi$, solving \rf{thridansatzeq}, for some values of $\alpha$, as well as the same functions for the usual BPS case, given in  \rf{ordinarybpssol}. In Figure \ref{fig:thirdansatz2} we plot the eigenvalues $\lambda_a$, $a=1,2$, defined in \rf{htausigmadiag2}, for solutions of the  equations \rf{thridansatzeq}, for some values of $\alpha$.

\subsection{The Wilson line}

We now evaluate the Wilson line, defined in \rf{weqdef}, for any gauge connection belonging to the 't Hooft-Polyakov radial ansatz \rf{thooftansatz}. We evaluate it on  curves, all starting at the same reference point, and divided into 3 parts, as follows. Consider a sphere a radius $R$, which will be taken to infinite at the end. The first part of the curve starts at the intersection of such a sphere with the $x^3$-axis, and slides on the sphere on the  $x^1\,x^3$-plane, up to an angle $\theta$. The second part slides on the sphere, from the end point of the first part, on a arc parallel to the $x^1\,x^2$-plane, up to angle $\phi$ with the $x^1\,x^3$-plane. Then the third part leaves the sphere on a radial direction towards the origin and stops at a distance $r$ from it. After the limit $R\rightarrow \infty$ is taken, any point $\(r\,,\, \theta\,,\, \phi\)$ of $\IR^3$ can be reached, from the reference point at the north pole of that infinite radius sphere, by a unique curve of such a family of curves. The parameterization is the following:\\
\noindent {\bf Part I:}
\br
x^1&=&R\,\sin \sigma
\nonumber\\
x^2&=&0\qquad\qquad\qquad\qquad 0\leq \sigma\leq \theta
\nonumber\\
x^3&=&R\,\cos \sigma
\nonumber
\er
\noindent {\bf Part II:}
\br
x^1&=&R\,\sin \theta\,\cos\(\sigma-\theta\)
\nonumber\\
x^2&=&R\,\sin \theta\,\sin\(\sigma-\theta\)\qquad\quad \theta\leq \sigma\leq \theta+\phi
\nonumber\\
x^3&=&R\,\cos \theta
\nonumber
\er
\noindent {\bf Part III:}
\br
x^1&=&\left[R-\(\sigma-\theta-\phi\)\,\(R-r\)\right]\,\sin \theta\,\cos\phi
\nonumber\\
x^2&=&\left[R-\(\sigma-\theta-\phi\)\,\(R-r\)\right]\,\sin \theta\,\sin\phi 
\nonumber\\
x^3&=&\left[R-\(\sigma-\theta-\phi\)\,\(R-r\)\right]\,\cos \theta
\nonumber
\er
with $\theta+\phi\leq \sigma\leq \theta+\phi+1$. 

The Wilson line is given by $W=W_{III}\,W_{II}\,W_I$ where $W_a$, $a=I,II,III$, is obtained by integrating \rf{weqdef} on each part $I$, $II$ and $III$. 

On part $I$ we have that ${\widehat r}_2=0$, since it is on the $x^1\,x^3$-plane, and  ${\widehat r}_3=\cos\sigma$ and ${\widehat r}_1=\sin\sigma$. Therefore
\be
A_i\,\frac{d\,x^i}{d\,\sigma}=\frac{1}{e}\,\(1-K\(R\)\)\,T_2
\lab{partonea}
\ee
and so
\be
W_{I}=e^{-i\,\(1-K\(R\)\)\,\theta\,T_2}
\ee
On part $II$ we have
\br
A_i\,\frac{d\,x^i}{d\,\sigma}&=&-\frac{1}{e}\,\(1-K\(R\)\)\,\sin\theta\times 
\\
&\times& e^{-i\,\(\sigma-\theta\)\,T_3}\,e^{-i\,\theta\,T_2}\,T_1\,e^{i\,\theta\,T_2}\,e^{i\,\(\sigma-\theta\)\,T_3}
\nonumber
\er
We then perform the gauge transformation $A_i\rightarrow{\bar A}_i=g\,A_i\,g^{-1}+\frac{i}{e}\,\partial_i g\,g^{-1}$, with $g=e^{i\,\theta\,T_2}\,e^{i\,\(\sigma-\theta\)\,T_3}$, to get
\be
{\bar A}_i\,\frac{d\,x^i}{d\,\sigma}=\frac{1}{e}\,\left[K\(R\)\,\sin \theta\,T_1-\cos\theta\,T_3\right]
\ee
Therefore 
\be
{\bar W}_{II}=e^{-i\,\phi\,\left[K\(R\)\,\sin \theta\,T_1-\cos\theta\,T_3\right]}
\ee
and so
\be
W_{II}=e^{-i\,\phi\,T_3}\,e^{-i\,\theta\,T_2}\,e^{-i\,\phi\,\left[K\(R\)\,\sin \theta\,T_1-\cos\theta\,T_3\right]}\,e^{i\,\theta\,T_2}
\ee
But that can be written as 
\be
W_{II}=e^{-i\,\phi\,\left[K\(R\)\,\sin \theta\,v_a^{(1)}\,T_a-\cos\theta\,{\widehat r}_a\,T_a\right]}\,e^{-i\,\phi\,T_3}
\ee
with $v_a^{(1)}$ given in \rf{zereigenvectors}. 

On part $III$ the line is along the radial direction, and since the radial part of the connection \rf{thooftansatz} vanishes, we have
\be
W_{III}=\one
\ee
The physically interesting field configurations are those where the profile function $K$ satisfies the boundary condition $K\(R\)\rightarrow 0$, as $R\rightarrow \infty$. Therefore, we get that  
\be
W=W_{III}\,W_{II}\,W_I= e^{i\,\phi\,\cos\theta\,{\widehat r}_a\,T_a}\,e^{-i\,\phi\,T_3}\,
e^{-i\,\theta\,T_2}
\ee
The adjoint matrix for such Wilson line is
\br
&&d\(W\)=
\lab{adjointw}\\
&&\(\cos\alpha\; v^{(1)}-\sin\alpha\;v^{(2)}\,,\,\sin\alpha\; v^{(1)}+\cos\alpha\;v^{(2)}\,,\,v^{(3)}\)
\nonumber
\er
with $\alpha=\phi\,\cos\theta$, and $v^{(a)}$, $a=1,2,3$, given in \rf{zereigenvectors}. But that is an orthogonal matrix that diagonalizes the matrix $h$, as in \rf{htausigmadiagonal}. Therefore, we indeed have that $M=d\(W\)$, and so the quantities $\({\cal F}_{\mu\nu}^a\,,\, \({\cal D}_{\mu}\Phi\)_a\)$,  and $\(\(F^W_{\mu\nu}\)^a\,,\, \left[ \(D_{\mu}\Phi\)^W\right]^a\)$, given respectively in \rf{fcaldef} and \rf{conjugateddef2}, coincide for the field configurations in the 't Hooft-Polyakov radial ansatz \rf{thooftansatz}. 

Another way of getting such results is to calculate the covariant derivatives of the vectors $v^{(a)}$, $a=1,2,3$, given in \rf{v3def} and \rf{zereigenvectors}, i.e. $D_{i}v^{(a)}=\partial_{i}v^{(a)}+i\,e\,d\(A_{i}\)\cdot v^{(a)}$, with $A_i$ given in \rf{thooftansatz}. One can check that
\be
D_r v^{(a)}\qquad\qquad a=1,2,3
\ee
and
\br
D_{\theta}v^{(1)}&=&-K(r)\,v^{(3)}
\nonumber\\
 D_{\theta}v^{(2)}&=&0
\\
D_{\theta}v^{(3)}&=&K(r)\,v^{(1)}
\nonumber
\er
and
\br
D_{\phi}v^{(1)}&=&\cos\theta\,v^{(2)}
\nonumber\\
 D_{\phi}v^{(2)}&=&-\cos\theta\,v^{(1)}-K(r)\,\sin\theta\,v^{(3)}
 \\
D_{\phi}v^{(3)}&=&K(r)\,\sin\theta \, v^{(2)}
\nonumber
\er
We can have $v^{(3)}$ covariantly constant, in all three directions,  at spatial infinity since we are assuming the boundary condition $K(r)\rightarrow 0$ as $r\rightarrow \infty$. Since $v^{(1)}$ and $v^{(2)}$ span the degenerate subspace we can take linear combinations of them. So, assuming $K\(\infty\)=0$, we impose that $D_{\phi}\left[F\, v^{(1)}+G\,v^{(2)}\right]=\left[\partial_{\phi}F-\cos\theta\,G\right]\,v^{(1)}+\left[\partial_{\phi}G+\cos\theta\,F\right]\,v^{(2)}=0$. Therefore, $\partial_{\phi}^2F+\cos^2\theta\,F=0$ and $\partial_{\phi}^2G+\cos^2\theta\,G=0$. So, $F$ and $G$ have to be sine and/or cossine of $(\phi\,\cos\theta)$. Taking
\br
{\widehat v}^{(1)}=\cos\(\phi\,\cos\theta\)\, v^{(1)}-\sin\(\phi\,\cos\theta\)\, v^{(2)}
\nonumber\\
{\widehat v}^{(2)}=\sin\(\phi\,\cos\theta\)\, v^{(1)}+\cos\(\phi\,\cos\theta\)\, v^{(2)}
\er
we get, at spatial infinity, 
\be
 D_{\theta}{\widehat v}^{(a)}=0\quad {\rm for} \quad\phi=0;\quad D_{\phi}{\widehat v}^{(a)}=0; \quad a=1,2
\ee
Therefore, the matrix $M=\({\widehat v}^{(1)}\,,\,{\widehat v}^{(2)}\,,\, v^{(3)}\)$, which is the same as \rf{adjointw}, satisfies the same equation as $W$, given in \rf{weqdef} (see \rf{covariantdermcurve}), on the curves described above \rf{partonea}. 

\section{Toroidal Solutions}
\label{sec:toroidal}
\setcounter{equation}{0}

We now construct an ans\"atz based on the three dimensional conformal symmetry of the model, discussed in appendix \ref{app:conformal}. Given an infinitesimal space transformation $x^i\rightarrow x^i+\zeta^i$, we say it is a symmetry of the equations of motion, if $A\(x\)\equiv A_i\(x\)\,dx^i$ and $\Phi\(x\)$ are solutions, then ${\tilde A}\(x\)=A\(x-\zeta\)$ and ${\tilde \Phi}\(x\)=\Phi\(x-\zeta\)$ are also solutions. Therefore
\br
{\tilde A}\(x\)&=&\left[A_i\(x\)-\zeta^j\,\partial_j A_i\(x\)\right]\left[dx^i-\partial_j \zeta^i\, dx^j\right]
\\
&=&A\(x\)-\left[\zeta^j\,\partial_j A_i\(x\)+\partial_i \zeta^j\,A_j\(x\)\right] dx^i+O\(\zeta^2\)
\nonumber
\er
and so, the variation of the fields are
\be
\delta A_i=-\zeta^j\,\partial_j A_i\(x\)-\partial_i \zeta^j\,A_j\(x\)\;;\qquad\delta\Phi=-\zeta^j\,\partial_j\Phi
\lab{variationconf}
\ee
Following \cite{babelon} we shall consider two commuting $U(1)$ subgroups of the conformal group corresponding to the vector fields, $V_{\zeta}=V_{\zeta^i}\partial_i$, given by
\br
\partial_{\phi}&\equiv& V_{\phi}= x_2\partial_1-x_1\partial_2
\lab{vectorfields}\\
\partial_{\xi}&\equiv& V_{\xi}=\frac{x_3}{a}\(x_1\partial_1+x_2\partial_2\)+\frac{1}{2\,a}\(a^2+x_3^2-x_1^2-x_2^2\)\partial_3
\nonumber
\er
where $a$ is an arbitrary length scale factor.  Note that we have introduced two angles $\phi$ and $\xi$, with translations along $\phi$ corresponding to rotations on the plane $x_1\,x_2$. The vector field $V_{\xi}$ is a linear combination of the special conformal transformation $x_3\,x_i\partial_i-\frac{1}{2}\,x_j^2\partial_3$, and the translation $\partial_3$. One can check that they indeed commute, i.e. $\sbr{\partial_{\phi}}{\partial_{\xi}}=0$. One can use such angles as coordinates on $\IR^3$, and complete the system with a third coordinate $z$, orthogonal to them, i.e. $\partial_{\phi}z=\partial_{\xi}z=0$. It turns out that those are the toroidal coordinates given by
\be
x_1=\frac{a}{p}\sqrt{z}\cos\phi\;;\;\; x_2=\frac{a}{p}\sqrt{z}\sin\phi\;;\;\; x_3=\frac{a}{p}\sqrt{1-z}\sin\xi
\lab{torocoord}
\ee
with $p=1-\sqrt{1-z}\,\cos\xi$, and $0\leq z\leq 1$, $0\leq \phi\,,\,\xi\leq 2\,\pi$. The metric is
\be
ds^2=\frac{a^2}{p^2}\left[ \frac{dz^2}{4\,z\(1-z\)}+\(1-z\)\,d\xi^2+z^2\,d\phi^2\right]
\lab{torometric}
\ee

There are some subtleties about the toroidal coordinates that are worth pointing. Note that
\br
r^2&=&x_1^2+x_2^2+x_3^2=a^2\,\frac{\(1+\sqrt{1-z}\,\cos\xi\)}{\(1-\sqrt{1-z}\,\cos\xi\)}
\nonumber\\
p&=&\frac{2}{1+r^2/a^2}
\lab{toror}
\er
and so, the spatial infinity corresponds to $z=0$ and $\xi=0$ (or $2\,\pi$). In addition, for $z=0$ the  angle $\phi$ looses its meaning, and so the toroidal coordinates contract all points on the two sphere $S^2_{\infty}$, at spatial infinity, to just one point. Consequently, it is perhaps correct to say that they are coordinates on the three sphere $S^3$ instead of $\IR^3$. That has consequences in what follows.

We shall consider two ans\"atze  based on the conformal symmetry of our system. The first requires that the solutions are invariant under the two commuting vector fields \rf{vectorfields}. So, taking $\zeta^i$ to be $\(0\,,\,0\,,\,\ve_{\phi}\)$, and $\(0\,,\,\ve_{\xi}\,,\,0\)$, respectively, with $\ve_{\phi}$ and $\ve_{\xi}$ constants, we get from \rf{variationconf} that the fields should not depend upon $\phi$ and $\xi$, i.e.
\be
A_i= {\hat A}_i^a\(z\)\,T_a\;;\qquad\quad \Phi={\hat \Phi}^a\(z\)\,T_a
\lab{firstansatz}
\ee
with $T_a$ being the generators of the gauge group. 

For the second ans\"atz we shall require the solutions to be invariant under the joint action of the two commuting vector fields \rf{vectorfields} and a gauge transformation, i.e. $A_i\rightarrow g\, A_i\,g^{-1}+\frac{i}{e}\,\partial_i g\,g^{-1}$, and $\Phi\rightarrow g\, \Phi\, g^{-1}$. Taking $g$ to be infinitesimally close to the identity element of the group, i.e. $g\sim\one+i\,\eta$, we get that $\delta A_i=-\frac{1}{e}\,D_i\eta$, with $D_i=\partial_i+i\,e\,\sbr{A_i}{}$, and $\delta \Phi=i\,\sbr{\eta}{\Phi}$. We have to choose two commuting $U(1)$ subgroups in the gauge group to compensate the action of the two commuting vector fields \rf{vectorfields}, generating two commuting $U(1)$ subgroups in the conformal group. We shall consider the case of $G=SU(2)$, where we can have at most one (commuting) $U(1)$ subgroup. So, taking $\zeta^i$ to be $\(0\,,\,0\,,\,\ve_{\phi}\)$, and $\eta=\ve_{\phi}\,n_{\phi}\,T_3$, with $\ve_{\phi}$ constant, we get that the invariance of the solutions under the joint action of such $U(1)$'s require that
\be
\partial_{\phi}A_i=i\, n_{\phi}\,\sbr{T_3}{A_i}\;;\qquad \partial_{\phi}\Phi=i\, n_{\phi}\,\sbr{T_3}{\Phi}
\ee
Similarly, taking $\zeta^i$ to be $\(0\,,\,\ve_{\xi}\,,\,0\)$, and $\eta=\ve_{\xi}\,n_{\xi}\,T_3$, with $\ve_{\xi}$ constant, the invariance of the solutions require
\be
\partial_{\xi}A_i=i\, n_{\xi}\,\sbr{T_3}{A_i}\;;\qquad \partial_{\xi}\Phi=i\, n_{\xi}\,\sbr{T_3}{\Phi}
\ee
The solutions satisfying those condition have the form
\br
A_i&=&{\tilde A}_i^3\(z\)\, T_3+{\tilde A}_i^+\(z\)\,e^{i\(n_{\xi}\,\xi+n_{\phi}\,\phi\)}\,T_{+}
\nonumber\\
&+&\({\tilde A}_i^+\(z\)\)^*\,e^{-i\(n_{\xi}\,\xi+n_{\phi}\,\phi\)}\,T_{-}
\nonumber\\
\Phi&=&{\tilde \Phi}^3\(z\)\, T_3+{\tilde \Phi}^+\(z\)\,e^{i\(n_{\xi}\,\xi+n_{\phi}\,\phi\)}\,T_{+}
\nonumber\\
&+&\({\tilde \Phi}^+\(z\)\)^*\,e^{-i\(n_{\xi}\,\xi+n_{\phi}\,\phi\)}\,T_{-}
\lab{secondansatz1}
\er
with $T_{\pm}=T_1\pm i\,T_2$, with $T_a$, $a=1,2,3$, being the generators of $SU(2)$, i.e. $\sbr{T_a}{T_b}=i\,\ve_{abc}\,T_c$. In order for the fields to be single valued we need $n_{\xi}$ and $n_{\phi}$ to be integers. In addition, note that $z=1$ corresponds to the circle of radius $a$, on the plane $x_1\,x_2$, and the angle $\xi$ looses its meaning there. Also, $z=0$ corresponds to the $x_3$-axis plus the spatial infinity, and the angle $\phi$ looses its meaning there. Therefore, for the solution to be single valued we need that
\be
{\tilde A}_i^+\(0\)={\tilde A}_i^+\(1\)=0\;;\quad  {\tilde \Phi}^+\(0\)={\tilde \Phi}^+\(1\)=0
\lab{singlecond}
\ee
Note that by performing a gauge transformation with $g=e^{-i\(n_{\xi}\,\xi+n_{\phi}\,\phi\)\,T_3}$, the fields \rf{secondansatz1} become
\br
A_{\xi}&=&\left[{\tilde A}_{\xi}^3\(z\)+\frac{n_{\xi}}{e}\right]\, T_3+{\tilde A}_{\xi}^1\(z\)\,T_{1}+{\tilde A}_{\xi}^2\(z\)\,T_{2}
\nonumber\\
A_{\phi}&=&\left[{\tilde A}_{\phi}^3\(z\)+\frac{n_{\phi}}{e}\right]\, T_3+{\tilde A}_{\phi}^1\(z\)\,T_{1}+{\tilde A}_{\phi}^2\(z\)\,T_{2}
\nonumber\\
A_{z}&=&{\tilde A}_z^a\(z\)\,T_a
\lab{secondansatz2}\\
\Phi&=&{\tilde \Phi}^a\(z\)\, T_a
\nonumber
\er
where we have denoted ${\tilde A}_i^+\(z\)=\({\tilde A}_i^1\(z\)-i\,{\tilde A}_i^2\(z\)\)/2$, and ${\tilde \Phi}^+\(z\)=\({\tilde \Phi}^1\(z\)-i\,{\tilde \Phi}^2\(z\)\)/2$.

Therefore, the ans\"atze \rf{firstansatz} and \rf{secondansatz2} are essentially the same, except that functions of the ans\"atz \rf{secondansatz2} are subjected to the condition \rf{singlecond}. Note in addition that if we take the $z$-component of the gauge potential to vanish, then gauge transformations with group elements of the form $g=e^{-i\(n_{\xi}\,\xi+n_{\phi}\,\phi\)\,T_3}$, keep that component zero. Therefore, we shall work with the ans\"atz \rf{firstansatz}, which is not subjected to conditions of the form \rf{singlecond}, with a vanishing $z$-component of the gauge potential, (dropping the hat from the notation of \rf{firstansatz})
\br
A_z&=&0\;;\qquad\qquad\quad\; A_{\xi}= A_{\xi}^a\(z\)\,T_a
\nonumber\\
 A_{\phi}&=& A_{\phi}^a\(z\)\,T_a\;;\qquad \Phi=\Phi^a\(z\)\,T_a
\lab{finalansatz}
\er
The field tensor is then given by
\be
F_{z\,\xi}=\partial_z A_{\xi}\;;\quad F_{z\,\phi}=\partial_z A_{\phi}\;;\quad F_{\xi\,\phi}=i\,e\,\sbr{A_{\xi}}{A_{\phi}}
\lab{toroidalfieldtensor}
\ee
and the covariant derivatives of the Higgs field are
\be
D_z\Phi=\partial_z\Phi\;;\;\;  D_{\xi}\Phi=i\,e\,\sbr{A_{\xi}}{\Phi}\;;\;\; D_{\phi}\Phi=i\,e\,\sbr{A_{\phi}}{\Phi}
\ee
As we commented above \rf{toror}, the spatial infinity corresponds to $z=0$ and $\xi=0$. Therefore, the solutions in the ans\"atz \rf{finalansatz} are constant on the two sphere $S^2_{\infty}$, at spatial infinity, as well as on the $x_3$-axis, since they do not depend upon $\xi$. That means that the topological magnetic charge \rf{magtopcharge} vanishes for all such solutions. Indeed, denoting $\left[r^2\,\trace\( B_i\, \Phi\)\right]_{z\rightarrow 0}\equiv c_i ={\rm constant}$, one gets
\br
\int_{S^2_{\infty}} d\Sigma_i\, \trace\( B_i\, \Phi\)&=&\int_0^{\pi}d\theta\,\int_0^{2\pi}d\phi\,\sin\theta\left[c_1\,\sin\theta\,\cos\phi
\right.\nonumber\\
&+&\left. c_2\,\sin\theta\,\sin\phi+c_3\,\cos\theta\right]=0
\lab{zerofluxtoro}
\er
However, we have used the Gauss theorem in \rf{magtopcharge}, and the Bianchi identity to write the topological charge as in \rf{topchargeymh}. So, if our solutions respect that theorem and identity, then \rf{topchargeymh} must also vanish. We then have ($\zeta^i=\(z\,,\,\xi\,,\,\phi\)$, and $\ve_{z\xi\phi}=1$)
\br
&&\int_{\IR^3} d^3x\,\trace\(B_i\,\(D_i\Phi\)\)=
\nonumber\\
&&-\frac{1}{2}\,\int_0^1dz\,\int_0^{2\pi}d\xi\,\int_0^{2\pi}d\phi\,\ve_{\zeta^i\zeta^j\zeta^k}\,\trace\(F_{\zeta^i\zeta^j}\,D_{\zeta^k}\Phi\)
\nonumber\\
&=&-ie4\pi^2\int_0^1dz\trace\left(\partial_zA_{\xi}\sbr{A_{\phi}}{\Phi}-\partial_zA_{\phi}\sbr{A_{\xi}}{\Phi}
\right.\nonumber\\
&+&\left. \sbr{A_{\xi}}{A_{\phi}}\partial_z\Phi\right)
\nonumber\\
&=&-ie4\pi^2\int_0^1dz\,\partial_z\trace\left(\sbr{A_{\xi}}{A_{\phi}}\,\Phi\right)
\er
Therefore the solutions have to satisfy
\be
\trace\left[\sbr{A_{\xi}}{A_{\phi}}\,\Phi\right]_{z=1}=\trace\left[\sbr{A_{\xi}}{A_{\phi}}\,\Phi\right]_{z=0}
\lab{torocond}
\ee

Denoting $B\equiv B_i\,dx^i=B_z\,dz+B_{\xi}\,d\xi+B_{\phi}\,d\phi$, one gets, from \rf{bfdef} and \rf{toroidalfieldtensor}, that
\br
B_z&=&-\frac{p}{a}\, \frac{i\,e}{2\,z\(1-z\)}\,\sbr{A_{\xi}}{A_{\phi}}
\nonumber\\
 B_{\xi}&=&2\,\frac{p}{a}\,\(1-z\)\,\partial_zA_{\phi}
\lab{toromagneticfield}\\
 B_{\phi}&=&-2\,\frac{p}{a}\,z\,\partial_zA_{\xi}
\nonumber
\er
Therefore, for the ans\"atz \rf{finalansatz} the self-duality equations  \rf{sdeqsymh} become
\br
\frac{e\,\ve_{bcd}}{2\,z\(1-z\)}\,A_{\xi}^c\(z\)\,A_{\phi}^d\(z\)\,{\widehat h}_{ba}\(z\)&=&\eta\,\partial_z\Phi^a\(z\)
\lab{finaltoroselfdualeqs}\\
2\,\(1-z\)\,\partial_zA_{\phi}^b\(z\)\,{\widehat h}_{ba}\(z\)&=&-\eta\,e\,\ve_{acd}\,A_{\xi}^c\(z\)\,\Phi^d\(z\)
\nonumber\\
2\,z\,\partial_zA_{\xi}^b\(z\)\,{\widehat h}_{ba}\(z\)&=&\eta\,e\,\ve_{acd}\,A_{\phi}^c\(z\)\,\Phi^d\(z\)
\nonumber
\er
where we have introduce the matrix ${\widehat h}_{ab}$ as
\be
h_{ab}\(z\,,\,\xi\)=\frac{a}{p}\,{\widehat h}_{ab}\(z\)
\lab{hhhatrel}
\ee
As we have argued, the self-dual solutions in the ans\"atz \rf{finalansatz}, satisfying \rf{torocond}, have zero topological charge, and so from \rf{energytopchargerel}, zero static energy. Therefore, if the eigenvalues of $h$ are all positive, we have that the static energy \rf{energyymh} is positive definite, and so the only possibility is that such solutions are trivial, i.e. $B_i=0$ and $D_i\Phi=0$. However, we now show that it is possible to have non-trivial self-dual solutions, with vanishing topological and static energy, but with the eigenvalues of the matrix $h$ not all positive. Such self-dual solutions are vacua solutions with non vanishing magnetic and Higgs fields. 

\subsection{A quasi-abelian solution}

Within the ans\"atz \rf{finalansatz} let us take
\be
A_{\xi}=\frac{1}{e}\,I\(z\)\,T_3\;;\qquad A_{\phi}=\frac{1}{e}\,J\(z\)\,T_3
\ee
and so, the condition \rf{torocond} is trivially satisfied. Then the first equation in \rf{finaltoroselfdualeqs} implies that the Higgs field must be constant, i.e.
\be
\Phi=\frac{1}{e}\,\gamma_a\, T_a\;;\qquad \gamma_a={\rm constant}
\lab{phiconstant}
\ee
The other two equations in \rf{finaltoroselfdualeqs} lead to (primes denote $z$-derivatives)
\be
2\(1-z\)\,\frac{J^{\prime}}{I}=-2\,z\,\frac{I^{\prime}}{J}=\eta\,\frac{\gamma_2}{{\widehat h}_{13}}=-\eta\,\frac{\gamma_1}{{\widehat h}_{23}}\;;\qquad
{\widehat h}_{33}=0
\ee
and the components ${\widehat h}_{11}$, ${\widehat h}_{22}$ and ${\widehat h}_{12}$, as well as the constant $\gamma_3$, are not constrained by the self-duality equations \rf{finaltoroselfdualeqs}. Such relations can be solved algebraically, without any integration, by taking
\be
I=-m_1\,\left[1-g\(z\)\right]\;;\qquad J=m_2\,g\(z\)
\ee
and leading to
\be
g=\frac{m_1^2\,z}{m_1^2\,z+m_2^2\,\(1-z\)}
\ee
and
\br
{\widehat h}_{13}&=&-\gamma_2\,f\;;\qquad\qquad {\widehat h}_{23}=\gamma_1\,f
\nonumber\\
f&=&\frac{\eta}{2\,m_1\,m_2}\,\left[m_1^2\,z+m_2^2\,\(1-z\)\right]
\er
The matrix ${\widehat h}$, defined in \rf{hhhatrel}, and its inverse are given by
\br
{\widehat h}&=&\left(
\begin{array}{ccc}
 {\widehat h}_{11} & {\widehat h}_{12} & -\gamma_2 f \\
 {\widehat h}_{12} & {\widehat h}_{22} & \gamma_1 f \\
 -\gamma_2 f & \gamma_1 f & 0 \\
\end{array}
\right)
\lab{hhatinvabelian}\\
{\widehat h}^{-1}&=&\frac{1}{\vartheta}\,\left(
\begin{array}{ccc}
 \gamma_1^2 & \gamma_1 \gamma_2 & -\frac{\gamma_1 {\widehat h}_{12}+\gamma_2 {\widehat h}_{22}}{f}
   \\
 \gamma_1 \gamma_2 & \gamma_2^2 & \frac{\gamma_1 {\widehat h}_{11}+\gamma_2 {\widehat h}_{12}}{f}
   \\
 -\frac{\gamma_1 {\widehat h}_{12}+\gamma_2 {\widehat h}_{22}}{f} & \frac{\gamma_1
   {\widehat h}_{11}+\gamma_2 {\widehat h}_{12}}{f} & \frac{{\widehat h}_{12}^2-{\widehat h}_{11} {\widehat h}_{22}}{f^2}
   \\
\end{array}
\right)
\nonumber
\er
where $\vartheta=\gamma_1^2 {\widehat h}_{11}+2 \gamma_1 \gamma_2 {\widehat h}_{12}+\gamma_2^2 {\widehat h}_{22}$. 

The gauge potential for such a solution is 
\br
A_z&=&0
\nonumber\\
A_{\xi}&=&-\frac{1}{e}\,\frac{m_1\,m_2^2\,\(1-z\)}{m_1^2\,z+m_2^2\,\(1-z\)}\,T_3
\lab{torogaugepot}\\
A_{\phi}&=&\frac{1}{e}\,\frac{m_2\,m_1^2\,z}{m_1^2\,z+m_2^2\,\(1-z\)}\,T_3
\nonumber
\er
From \rf{toromagneticfield} we get that the magnetic field is
\be
B_i=\alpha\,A_i\;;\qquad \alpha=-2\,\frac{p}{a}\,\frac{m_1\,m_2}{\left[m_1^2\,z+m_2^2\,\(1-z\)\right]}
\lab{bfinalabelian}
\ee
As we have seen, the spatial infinity corresponds to $z\rightarrow 0$ and $\xi\rightarrow 0$. Then, using \rf{toror}, one can check that $B_{\xi}\rightarrow 1/r^2$, and $B_{\phi}\rightarrow 1/r^4$, as $r\rightarrow \infty$. Despite the Coulomb like tail of the $\xi$-component of the magnetic field, the integrated magnetic flux on a two-sphere at spatial vanishes as argued in \rf{zerofluxtoro}. 

Note that we are working with the components of the one-forms, i.e. $A=A_i\,dx^i$ and $B=B_i\,dx^i$. If we work instead with the components of the vectors, in terms of the unit vectors of the coordinate system, i.e. ${\vec A}={\bar A}_i\,{\vec e}_i$ and ${\vec B}={\bar B}_i\,{\vec e}_i$, the relation above is kept unchanged, i.e. ${\vec B}=\alpha\, {\vec A}$, since both sides change the same way. We are working with abelian gauge fields and so the magnetic field is the curl of ${\vec A}$. Therefore,  the vector ${\vec A}$ is a force free field, i.e. ${\vec \nabla}\wedge {\vec A}=\alpha\,{\vec A}$, and the solution we have may be of interest in magnetohydrodynamics  \cite{marsh,bpsshnir}.

The components of the magnetic vector field in terms of the unit vector of the coordinate systems, i.e. ${\vec B}={\bar B}_i\,{\vec e}_i={\bar B}_{\zeta^i}\,{\vec e}_{\zeta^i}$, with $\(\zeta^1\,,\,\zeta^2\,,\,\zeta^2\)=\(z\,,\,\xi\,,\,\phi\)$, are given by
\br
{\bar B}_z&=&0
\nonumber\\
 {\bar B}_{\xi}&=&\frac{2}{e}\,\frac{p^2}{a^2}\,\frac{m_1^2\,m_2^3\,\sqrt{1-z}}{\left[m_1^2\,z+m_2^2\,\(1-z\)\right]^2}\,T_3
\lab{abelianmagneticvectorfield}\\
{\bar B}_{\phi}&=&-\frac{2}{e}\,\frac{p^2}{a^2}\,\frac{m_1^3\,m_2^2\,\sqrt{z}}{\left[m_1^2\,z+m_2^2\,\(1-z\)\right]^2}\,T_3
\nonumber
\er
Again, using \rf{toror}, one can check that 
${\bar B}_{\xi}\rightarrow 1/r^4$, and ${\bar B}_{\phi}\rightarrow 1/r^5$, as $r\rightarrow \infty$. 

\begin{figure*}
\centering
		\includegraphics[scale=1.45]{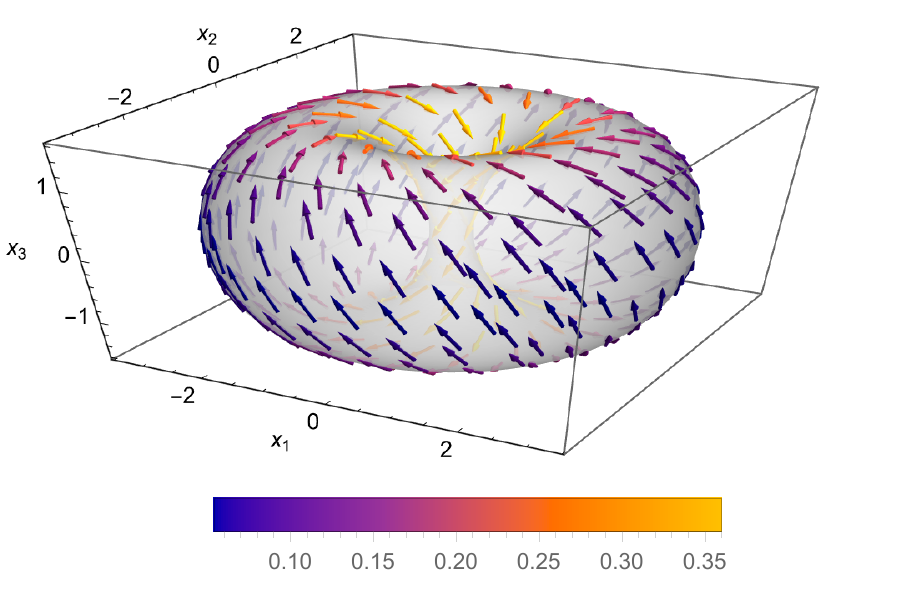}
		\caption{The  magnetic field vector  \rf{abelianmagneticvectorfield} for $m_1=1$ and $m_2=1$, and for $z=0.3$.  The colors refer to the modulus of the magnetic field.}
		\label{fig:m11m21}
\centering 
\end{figure*}

\begin{figure*}
\centering
		\includegraphics[scale=1.45]{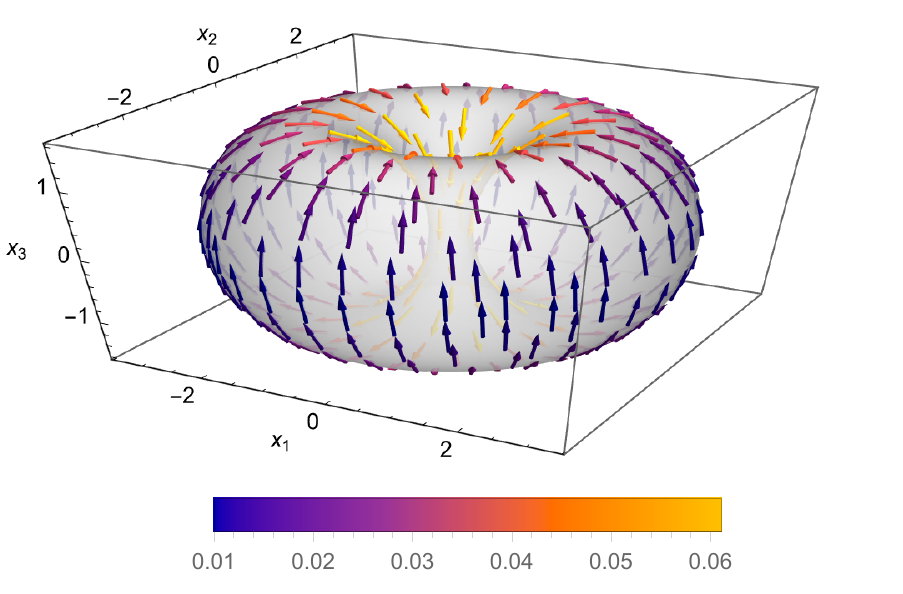}
		\caption{The  magnetic field vector  \rf{abelianmagneticvectorfield} for $m_1=1$ and $m_2=10$, and for $z=0.3$.  The colors refer to the modulus of the magnetic field.}
		\label{fig:m11m210}
\centering
\end{figure*}

\begin{figure*}
\centering
		\includegraphics[scale=1.45]{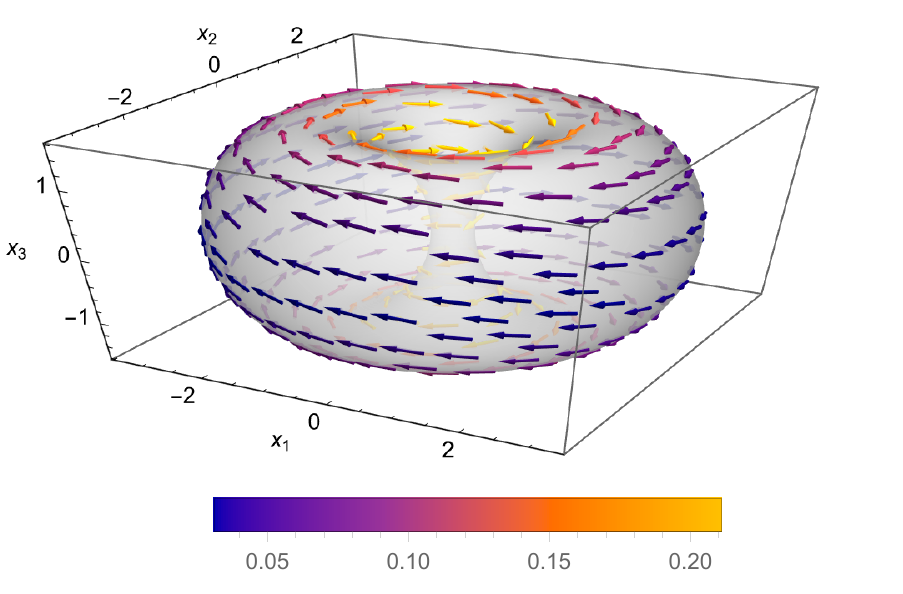}
		\caption{The  magnetic field vector  \rf{abelianmagneticvectorfield} for $m_1=10$ and $m_2=1$, and for $z=0.3$.  The colors refer to the modulus of the magnetic field.}
		\label{fig:m110m21}
\centering
\end{figure*}

In Figures \ref{fig:m11m21}, \ref{fig:m11m210} and \ref{fig:m110m21} we plot the magnetic vector \rf{abelianmagneticvectorfield} for the $\(m_1\,,\,m_2\)=\(1\,,\,1\)$, $\(m_1\,,\,m_2\)=\(1\,,\,10\)$ and $\(m_1\,,\,m_2\)=\(10\,,\,1\)$, respectively, for $z=0.3$. 

Note that we can take either $\gamma_1$ or $\gamma_2$ to vanish, but we can not take both to vanish, since the matrix $h$ would not be invertible. 

From \rf{tausigmadef}, \rf{torometric}, \rf{torogaugepot} and \rf{bfinalabelian}, one can check that all components of the matrix $\tau_{ab}$ vanish except for $\tau_{33}=\frac{\eta}{e^2}\frac{p^4}{a^4}\frac{m_1\,m_2}{2\,f^3}$. Therefore, the matrices $\tau$ and $h$ do not commute, and $\sigma$ is not symmetric. In fact, all components of the matrix $\sigma$ vanish except for 
$\sigma_{31}=-\frac{\gamma_2}{e^2}\frac{p^3}{a^3}\frac{m_1\,m_2}{2\,f^2}$ and $\sigma_{32}=\frac{\gamma_1}{e^2}\frac{p^3}{a^3}\frac{m_1\,m_2}{2\,f^2}$.

One can check, using \rf{phiconstant}, \rf{hhatinvabelian}, \rf{torogaugepot} and \rf{bfinalabelian}, that the two terms of the energy density in \rf{energyymh} vanish independently, i.e. $h_{ab}\,B_i^a\,B_i^b=0$ and $h^{-1}_{ab}\,\(D_i\Phi\)^a\,\(D_i\Phi\)^b=0$, and so the static energy of such a solution is indeed zero, as well as its topological charge \rf{topchargeymh}. 

However, such a solution does possess another topological charge  which is the winding number of the maps $S^3\rightarrow S_T^3$, where $S^3$ is $\IR^3$ with the spatial infinity identified to a point, and $S^3_T$ is the target three sphere parametrized by two complex fields $Z_a$, $a=1,2$, such that $\mid Z_1\mid^2+\mid Z_2\mid^2=1$. Let us now consider the following configurations of such fields as  
\be
Z_1=\sqrt{1-g\(z\)}\,e^{i\,m_1\,\xi}\;;\qquad Z_2=\sqrt{g\(z\)}\,e^{-i\,m_2\,\phi}
\ee
Consider the vector field
\be
{\cal A}_i=\frac{i}{2}\(Z_a^{\dagger}\partial_i Z_a-Z_a\partial_i Z_a^{\dagger}\)=i\,Z_a^{\dagger}\partial_i Z_a
\ee
One can check that
\be
{\cal A}_i=e\, \trace\(A_i\,T_3\)
\lab{identifya}
\ee
with $A_i$ given in \rf{torogaugepot}. The topological charge is given by the integral representation of the Hopf invariant, i.e. 
\be
Q_H=\frac{1}{4\,\pi^2}\int d^3x\,\ve_{ijk}\,{\cal A}_{i}\,\partial_j{\cal A}_k
\lab{hopfcharge}
\ee
However, we do not perform the projection of $S_T^3$ into $S_T^2$, as $\(Z_1\,,\, Z_2\)\rightarrow u\equiv Z_2/Z_1$, with $u$ parametrizing a complex plane which is the stereographic projection of $S_T^2$. Therefore, $Q_H$, given in \rf{hopfcharge}, is indeed the winding number of $S^3 \rightarrow S^3_T$, where $S^3$ is  $\IR^3$ with the spatial infinity identified to a point. Such an identification can be done because the solutions go to a constant at spatial infinity. 

Evaluating the topological charge \rf{hopfcharge} on the solutions \rf{torogaugepot} and \rf{identifya} one gets
\be
Q_H=m_1\,m_2
\ee
where we have used the fact that $d^3x\,\ve_{ijk}\,{\cal A}_{i}\,\partial_j{\cal A}_k=d^3\zeta\, \ve_{\zeta^i\zeta^j\zeta^k}\,{\cal A}_{\zeta^i}\partial_{\zeta^j}{\cal A}_{\zeta^k}$, with $\(\zeta^1\,,\,\zeta^2\,,\,\zeta^3\)=\(z\,,\,\xi\,,\,\phi\)$, and $\ve_{z\xi\phi}=1$.

Note that the solutions \rf{torogaugepot} and \rf{identifya} are the same as the ones obtained in \cite{bpsshnir} for a modified $SU(2)$ Skyrme model.

So, despite the fact that we have vacuum solutions with vanishing energy and magnetic charge, such solutions do present a non-trivial topological charge, given by \rf{hopfcharge}, and non-trivial toroidal magnetic fields. Note that even though the energy vanishes, its density does not, and so the energy can not be positive definite, and consequently the eigenvalues of the $h$-matrix can not be all positive. It would be interesting to investigate the stability of such solutions, and find if the non-trivial topological charge \rf{hopfcharge} may impose some selection rules. 

Note that the components ${\widehat h}_{11}$, ${\widehat h}_{12}$ and ${\widehat h}_{22}$ appearing in the matrix \rf{hhatinvabelian} were not fixed by the self-duality equations as functions of the gauge and Higgs fields. Therefore, the matrix $M$ that diagonalizes $h$, as in \rf{diaghgeneral}, will not depend only on the gauge fields, and consequently $M$ can not be related to the adjoint matrix of the Wilson line $W$. 

\subsection{A simple non-abelian solution}

Again within the ans\"atz \rf{finalansatz} let us take
\br
A_{\xi}&=&\frac{1}{e}\,\(1-z\)\,H_1\(z\)\,T_1
\nonumber\\
A_{\phi}&=&\frac{1}{e}\,z\,H_2\(z\)\,T_2
\lab{gaugehiggsnonabelian}\\
\Phi&=&\frac{1}{e}\,H_3\(z\)\,T_3
\nonumber
\er
and the condition \rf{torocond} leads to
\be
\left[z\(1-z\)\,H_1\,H_2\,H_3\right]_{z=0}=\left[z\(1-z\)\,H_1\,H_2\,H_3\right]_{z=1}
\lab{torocndnonabelian}
\ee
which is satisfied as long as the functions $H_a$, $a=1,2,3$, are finite at $z=0$ and at $z=1$. 

The self-duality equations \rf{finaltoroselfdualeqs} imply that the matrix $h$ is diagonal, i.e.
\be
{\widehat h}_{ab}=\lambda_a\(z\)\,\delta_{ab}
\lab{nonabelianhhat}
\ee
and its diagonal elements are completely determined in terms of the functions $H_a\(z\)$ as
\br
\lambda_1&=& \frac{\eta}{2}\,\frac{H_2\,H_3}{\left[\(1-z\)\,H^{\prime}_1-H_1\right]}
\nonumber\\
\lambda_2&=& \frac{\eta}{2}\,\frac{H_1\,H_3}{\left[ z\,H^{\prime}_2+H_2\right]}
\lab{eigenvaluesh}\\
\lambda_3&=& 2\,\eta\,\frac{H^{\prime}_3}{H_1\,H_2}
\nonumber
\er
The self-duality equations \rf{finaltoroselfdualeqs} do not impose any condition on the functions $H_a$. The only requirement on such functions is that none of the $\lambda_a$, $a=1,2,3$, given in \rf{eigenvaluesh}, can vanish identically, since that would imply that the matrix $h$ is not invertible. 

The magnetic field \rf{toromagneticfield} and the covariant derivative of the Higgs field become
\br
B_z&=&\frac{1}{2\,e}\,\frac{p}{a}\,H_1\,H_2\,T_3
\nonumber\\
B_{\xi}&=&\frac{2}{e}\,\frac{p}{a}\,\(1-z\)\,\left[ z\,H^{\prime}_2+H_2\right]\,T_2
\nonumber\\
B_{\phi}&=&-\frac{2}{e}\,\frac{p}{a}\,z\,\left[\(1-z\)\,H^{\prime}_1-H_1\right]\,T_1
\lab{toromagneticfieldnonabelian}\\
 D_z\Phi&=&\frac{1}{e}\,H_3^{\prime}\,T_3
 \nonumber\\
 D_{\xi}\Phi&=&\frac{1}{e}\,\(1-z\)\,H_1\,H_3\,T_2
 \nonumber\\
 D_{\phi}\Phi&=&-\frac{1}{e}\,z\,H_2\,H_3\,T_1
 \nonumber
\er
From \rf{tausigmadef}, \rf{torometric}, and \rf{toromagneticfieldnonabelian} one observes that, in this case,  the matrices $\tau$ and $\sigma$ are also diagonal.

Note that the eigenvalues \rf{eigenvaluesh} of $h$ can not  have all the same sign, if the condition \rf{torocond}, or equivalently \rf{torocndnonabelian}, is satisfied.  Indeed, if all the eigenvalues \rf{eigenvaluesh} of $h$ have the same sign, then it follows that $H_1\,H_2\,H_3^{\prime}$, $H_1\,\left[ z\,H^{\prime}_2+H_2\right]\,H_3$, and $\left[\(1-z\)\,H^{\prime}_1-H_1\right]\,H_2\,H_3$, all have the sign. Since $z$ and $\(1-z\)$ are positive, it follows that $\partial_z\left[\(1-z\)\,H_1\,z\,H_2\,H_3\right]$ is either strictly positive or strictly negative, and so its integral on the interval $z\in \left[0\,,\,1\right]$ can not vanish. But that contradicts the condition \rf{torocndnonabelian}.
 
\noindent One can check, using \rf{hhhatrel}, \rf{nonabelianhhat} and \rf{eigenvaluesh}, that 
$d^3x\,h_{ab}\,B_i^a\,B_i^b=d^3x\,h^{-1}_{ab}\,\(D_i\Phi\)^a\,\(D_i\Phi\)^b=\(\eta/e^2\)\,dz\,d\xi\,d\phi\,\partial_z\left[z\(1-z\)\,H_1\,H_2\,H_3\right]$. Therefore, the static energy \rf{energyymh} indeed vanishes for such solutions, if the functions $H_a$, $a=1,2,3$, are finite at $z=0$ and $z=1$, i.e. they satisfy \rf{torocond} or equivalently \rf{torocndnonabelian}. 

Using \rf{toror} and the fact that the spatial infinity corresponds to $z\rightarrow 0$ and $\xi\rightarrow 0$, one gets that, if the functions $H_a$, $a=1,2$ remain finite at $z=0$, then $B_z\rightarrow 1/r^2$, $ B_{\xi}\rightarrow 1/r^2$, and $B_{\phi}\rightarrow 1/r^4$, as $r\rightarrow \infty$. Despite the fact that the $z$ and $\xi$-components of the magnetic field present a Coulomb like tail, the magnetic flux, integrated over a two-sphere at infinity, vanishes.  The reason, as argued in \rf{zerofluxtoro}, is that since the magnetic field depends on $z$ and $\xi$ only, and since those have a fixed value at spatial infinity, namely $z=0$ and $\xi=0$, it has a constant direction in space, and in the algebra, and so the integrated flux vanishes.

\begin{figure*}
\centering 
		\includegraphics[scale=1.45]{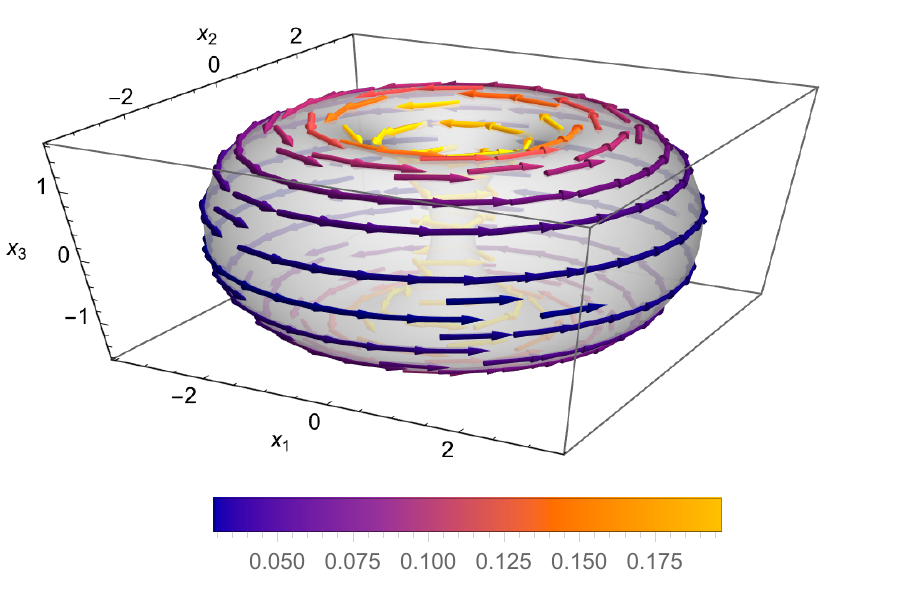}
		\caption{The component of magnetic field \rf{toromagneticfieldnonabelianvec} in the direction of the generator $T_1$ of the $SU(2)$ Lie algebra, for $z=0.3$, and $H_1=H_2=1$.  The colors refer to the modulus of the magnetic field.}
		\label{fig:toro1}
\centering 
\end{figure*}

\begin{figure*}
\centering
		\includegraphics[scale=1.45]{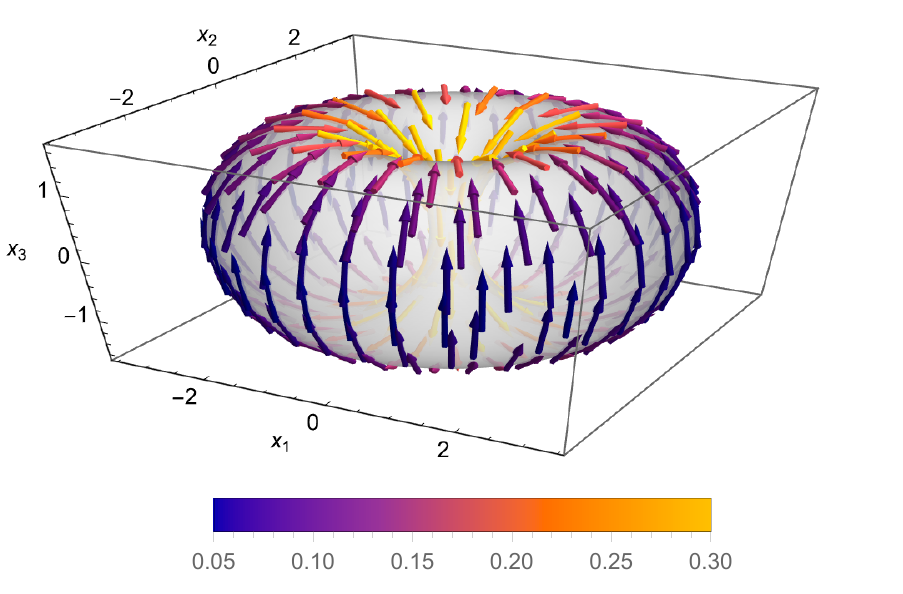}
		\caption{The component of magnetic field \rf{toromagneticfieldnonabelianvec} in the direction of the generator $T_2$ of the $SU(2)$ Lie algebra, for $z=0.3$, and $H_1=H_2=1$.  The colors refer to the modulus of the magnetic field.}
		\label{fig:toro2}
\centering 
\end{figure*}

\begin{figure*}
		\includegraphics[scale=1.45]{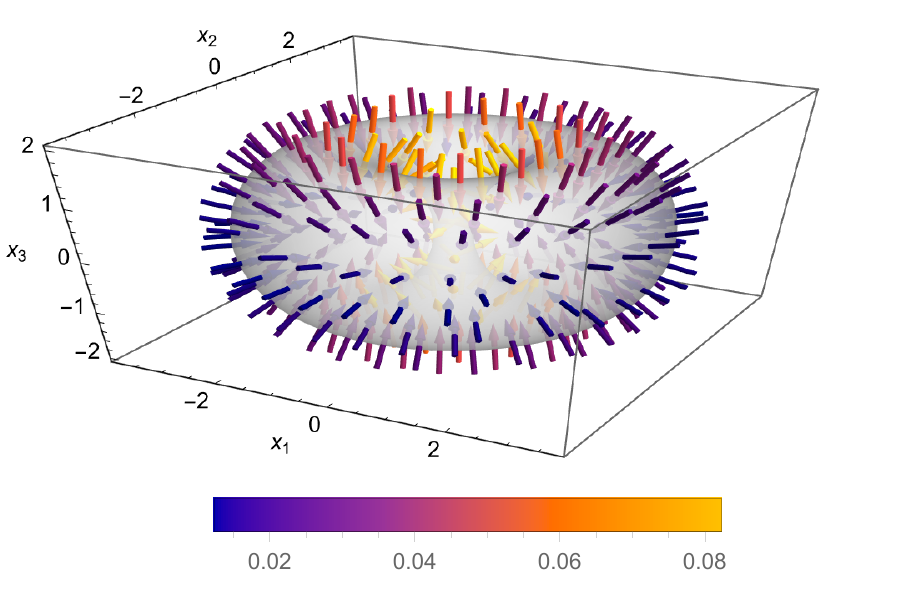}
		\caption{The component of magnetic field \rf{toromagneticfieldnonabelianvec} in the direction of the generator $T_3$ of the $SU(2)$ Lie algebra, for $z=0.3$, and $H_1=H_2=1$.  The colors refer to the modulus of the magnetic field.}
		\label{fig:toro3}
\centering
\end{figure*}

Note that the components of the magnetic field given in \rf{toromagneticfieldnonabelian} are the components of the one-form $B=B_i\,dx^i=B_{\zeta^i}\,d\zeta^i$, with $\(\zeta^1\,,\,\zeta^2\,,\,\zeta^2\)=\(z\,,\,\xi\,,\,\phi\)$. If we write the magnetic field vector in terms of the unit vectors of the coordinate systems, i.e., 
${\vec B}={\bar B}_i\,{\vec e}_i={\bar B}_{\zeta^i}\,{\vec e}_{\zeta^i}$, we get that
\br
{\bar B}_z&=&\frac{1}{e}\,\frac{p^2}{a^2}\,\sqrt{z\(1-z\)}\,H_1\,H_2\,T_3
\nonumber\\
{\bar B}_{\xi}&=&\frac{2}{e}\,\frac{p^2}{a^2}\,\sqrt{1-z}\,\left[ z\,H^{\prime}_2+H_2\right]\,T_2
\lab{toromagneticfieldnonabelianvec}\\
{\bar B}_{\phi}&=&-\frac{2}{e}\,\frac{p^2}{a^2}\,\sqrt{z}\,\left[\(1-z\)\,H^{\prime}_1-H_1\right]\,T_1
\nonumber
\er
Again using \rf{toror}, and if the functions $H_a$, $a=1,2$ remain finite at $z=0$, one gets that ${\bar B}_z\rightarrow 1/r^5$, ${\bar B}_{\xi}\rightarrow 1/r^4$, and ${\bar B}_{\phi}\rightarrow 1/r^5$, as $r\rightarrow \infty$. In Figures \ref{fig:toro1}, \ref{fig:toro2} and \ref{fig:toro3} we plot the components of the magnetic field vector \rf{toromagneticfieldnonabelianvec} in the direction of the generators $T_1$, $T_2$ and $T_3$, respectively, of the $SU(2)$ Lie algebra, for $z=0.3$, and $H_1=H_2=1$.

\subsubsection{The Wilson line}

We now evaluated the Wilson line for gauge connections belonging to the toroidal ansatz \rf{gaugehiggsnonabelian}. All the curves we consider start at the same fixed reference point and are divided in three parts. Consider a toroidal surface of thickness $z=z_0$, i.e. the surface obtained, through the toroidal coordinates \rf{torocoord}, by fixing the value of the coordinate $z$ to $z_0$, and varying  both angles $\xi$ and $\phi$ from $0$ to $2\,\pi$. The fixed reference point is the intersection of that toroidal surface with the $x_1$-axis. The first part of the curve starts at that reference point and slides on the toroidal surface on the $x_1\,x_2$-plane ($\xi=0$), on the anti-clockwise direction up to an angle $\phi$. The second part stars at the end  of the first part of the curve, sliding the toroidal surface upward (increasing $\xi$) keeping the  value of the angle $\phi$ fixed, up to and angle $\xi$. Then the third part starts at the end of the second part of the curve, leaving the toroidal surface, either upward or downward, keeping the values of the angles $\xi$ and $\phi$ fixed, up to a given value of the coordinate $z$. After we taking the limit $z_0\rightarrow 0$, or $z_0\rightarrow 1$, any point of $\IR^3$ can be reached by a unique curve of such a family of curves. The parameterization is the following:\\
\noindent {\bf Part I:}
\br
x^1&=&\frac{a}{p_I}\,\sqrt{z_0}\,\cos \sigma
\nonumber\\
x^2&=&\frac{a}{p_I}\,\sqrt{z_0}\,\sin \sigma\qquad\qquad\qquad 0\leq \sigma\leq \phi
\nonumber\\
x^3&=&0
\nonumber
\er
with $p_I=1-\sqrt{1-z_0}$.\\
\noindent {\bf Part II:}
\br
x^1&=&\frac{a}{p_{II}}\,\sqrt{z_0}\,\cos \phi
\nonumber\\
x^2&=&\frac{a}{p_{II}}\,\sqrt{z_0}\,\sin \phi\qquad\qquad\qquad \phi\leq \sigma\leq \phi+\xi
\nonumber\\
x^3&=&\frac{a}{p_{II}}\,\sqrt{1-z_0}\,\sin\(\sigma-\phi\)
\nonumber
\er
with $p_{II}=1-\sqrt{1-z_0}\cos\(\sigma-\phi\)$.\\
\noindent {\bf Part III:}
\br
x^1&=&\frac{a}{p_{III}}\,\sqrt{w\(\sigma\)}\,\cos \phi
\nonumber\\
x^2&=&\frac{a}{p_{III}}\,\sqrt{w\(\sigma\)}\,\sin \phi\qquad\qquad \phi+\xi\leq \sigma\leq \phi+\xi+1
\nonumber\\
x^3&=&\frac{a}{p_{III}}\,\sqrt{1-w\(\sigma\)}\,\sin\xi
\nonumber
\er
with $p_{III}=1-\sqrt{1-w\(\sigma\)}\cos\xi$, and $w\(\sigma\)=z_0-\(\sigma-\phi-\xi\)\(z_0-z\)$. Note that in Part III we can have either $z>z_0$ or $z<z_0$.

The Wilson line is given by $W=W_{III}\,W_{II}\,W_I$ where $W_a$, $a=I,II,III$, is obtained by integrating \rf{weqdef} on each part $I$, $II$ and $III$. The integration of \rf{weqdef} is quite simple because in Parts I, II and III the curves are along the $\phi$, $\xi$ and $z$ directions respectively, and so $A_i\,\frac{d\,x^i}{d\,\sigma}= A_{\zeta}$, with $\zeta$ equal to $\phi$, $\xi$ and $z$  respectively. But $A_{\phi}$ and $A_{\xi}$ depend only on $z$, and $A_z=0$. So we get
\br
W_I&=&e^{-i\,z_0\,H_2\(z_0\)\,\phi\,T_2}
\nonumber\\
W_{II}&=&e^{-i\,\(1-z_0\)\,H_1\(z_0\)\,\xi\,T_1}
\lab{w123}\\
W_{III}&=&\one
\nonumber
\er
We now consider configurations satisfying the boundary conditions $H_1\(0\)=0$ and $H_2\(1\)=0$, which are quite compatible with the condition \rf{torocndnonabelian}. Therefore, when we take the limit $z_0\rightarrow 0$ (infinitely thick torus), or $z_0\rightarrow 1$ (infinitesimally thin torus), we get that 
\be
W=\one
\ee
Since for the ansatz \rf{gaugehiggsnonabelian} the matrix $h$ is already diagonal (see \rf{nonabelianhhat}) we get that the matrix $M$ is unity and so we have in such a  case that $M=d\(W\)$. 

Another way of obtaining such result is to analyse the covariant derivatives of the matrix $M$, which  in this case is unity, i.e. $M=\one$. For the ansatz \rf{gaugehiggsnonabelian}, where $A_z=0$, we get $D_z\one=0$, and
\br
D_{\xi}\one= i\,\(1-z\)\,H_1\(z\)\,d\(T_1\);\quad D_{\phi}\one= i\,z\,H_2\(z\)\,d\(T_2\)
\nonumber
\er
By assuming the boundary conditions $H_1\(0\)=0$ and/or $H_2\(1\)=0$, one observes that $M=\one$ satisfies the same equation as $W$, given in \rf{weqdef} (see \rf{covariantdermcurve}), on the curves described above   \rf{w123}, for $z_0\rightarrow 0$ or $z_0\rightarrow 1$.

\section{Conclusions}
\label{sec:conclusions}
\setcounter{equation}{0}

We have explored the concept of generalized self-duality in the context of the Yang-Mills-Higgs system by the introduction of $N(N+1)/2$ scalar fields, where $N$ is the dimension of the gauge group $G$. Those fields are assembled in a symmetric and invertible matrix $h_{ab}$, that transforms under the symmetric part of the direct product of the adjoint representation of $G$ with itself. The coupling of such fields to the gauge and Higgs field is made by the replacement of the Killing form of $G$, in the contraction of group indices, by $h$ in the kinetic term of the gauge fields, and by its inverse in the Higgs fields kinetic term. The theory we consider does not present a Higgs potential, neither one in the Prasad-Sommerfield limit. 

The introduction of the $h$-fields renders our modified Yang-Mills-Higgs system conformally invariant in the three dimensional space $\IR^3$, bringing interesting new features to it. The generalized self-duality equations are such that, given a (perhaps any) configuration of the gauge and Higgs fields, the $h$-fields adjust themselves to solve those equations. So, our model possesses plenty of solutions. Indeed, we have constructed many solutions using the 't Hooft-Polyakov spherically symmetric ans\"atz in the case $G=SU(2)$, and also using the conformal symmetry to build toroidal ans\"atz to construct  vacuum configurations presenting non-trivial toroidal magnetic field configurations. 

The physical role of the $h$-fields is still far from clear, and new investigations are necessary to clarify that issue. We have shown however that by diagonalizing $h$, i.e. $h=M\,h_D\,M^T$, where $h_D$ is diagonal and $M$ an orthogonal matrix, it turns out that the $h_D$-fields play the role of dilaton fields leading to the conformal symmetry of the theory in the three dimensional space $\IR^3$. The $M$-fields relate, in many cases, to the Wilson line operator in the adjoint representation and lead to dressed quantities, like the field tensor and covariant derivative of the Higgs field, that become gauge invariant. Those facts points to an interpretation of the theory \rf{actionymh} as an effective Yang-Mills-Higgs theory. It would be interesting to study that further and explore its consequences.That would open up new ways of studying the Yang-Mills-Higgs system.

The special coupling of the $h$-fields to the gauge and Higgs fields, which leads to self-duality, did not allow the introduction of kinetic and potential terms for them. It would be interesting to investigate that route of breaking the self-duality, even in a perturbative way, and explore the physical consequences of it. The $h$-fields have been introduced in the Skyrme model, leading to an exact self-dual sector \cite{laf2017,us}, and they have lead to new applications of the Skyrme model to nuclear matter \cite{nuclear}. In fact, there may be a connection to be explored among magnetic monopoles of the Yang-Mills-Higgs system, presented here, and Skyrmions in the models \cite{laf2017,us}.

\vspace{1cm}

\noindent {\bf Acknowledgments:} LAF is partially supported by Conselho Nacional de Desenvolvimento Cient\'ifico e Tecnol\'ogico - CNPq (contract 308894/2018-9), and HM is supported by a scholarship from Funda\c c\~ao de Amparo \`a Pesquisa do Estado de S\~ao Paulo — FAPESP (contract 2018/21601-6).

\appendix

\section{Conformal Symmetry}
\label{app:conformal}
\setcounter{equation}{0}

We show in this appendix that the self-duality equations \rf{sdeqsymh} and the static energy \rf{energyymh} are conformally invariant in the three dimensional space $\IR^3$. We consider space transformations of the form 
\be
\delta x^i=\zeta^i
\ee
with the infinitesimal parameters $\zeta^i$ satisfying
\be
\partial_i\zeta_j+\partial_j\zeta_i=2\,\Omega\, \delta_{ij}
\ee
For spatial rotations and translations we have that $\Omega=0$, for dilatations we have that $\Omega$ is constant, and for special conformal transformations we have that $\Omega$ is linear in the Cartesian coordinates $x^i$. The fields transform as
\br
\delta\,A_{i}&=&-\partial_i\zeta^j\,A_j\;;\qquad \delta F_{ij}=-\partial_i\zeta^k\,F_{kj}-\partial_j\zeta^k\,F_{ik}
\nonumber\\
\delta D_i\Phi&=&-\partial_i\zeta^j\,D_j\Phi\qquad \delta h_{ab}=\Omega\,h_{ab}\;;\qquad 
\er
The magnetic field \rf{bfdef} transform as
\br
\delta B_i&=&\ve_{ijk}\,\partial_j\zeta_l\,F_{lk}
=-\ve_{ijk}\,\ve_{lkm}\partial_j\zeta_l\,B_m
\\
&=&\partial_j\zeta_i\,B_j-\partial_j\zeta_j\,B_i=\partial_j\zeta_i\,B_j-3\,\Omega\,B_i
\nonumber
\er
Therefore, we have that 
\br
\delta\(h_{ab}\,B_{i}^a\,B_{i}^b\)&=&-3\,\Omega\,h_{ab}\,B_{i}^a\,B_{i}^b
\\
\delta\(h^{-1}_{ab}\,\(D_i\Phi\)^a\, \(D_i\Phi\)^b\)&=&-3\,\Omega\,h^{-1}_{ab}\,\(D_i\Phi\)^a\, \(D_i\Phi\)^b
\nonumber
\er
Using the fact that the volume element transfom as $\delta\(d^3x\)=3\,\Omega\,d^3x$, we conclude that the static energy \rf{energyymh} is conformally invariant. Denoting the self-duality equations \rf{sdeqsymh} as
\be
{\cal E}_{ia}\equiv B_{i}^b\,h_{ba}-\eta\,\(D_i\Phi\)^a
\ee
one gets
\br
\delta {\cal E}_{ia}&=&\partial_j\zeta_i\,B_j^b\,h_{ba}-2\,\Omega\,B_i^b\,h_{ba}+\eta\,\partial_i\zeta_j\,\(D_j\Phi\)^a
\nonumber\\
&=&-\partial_i\zeta_j\,{\cal E}_{ja}
\er
Therefore, the self-duality equations are conformally invariant. One can check that the static Euler-Lagrange for the gauge, Higgs and $h$ fields are also conformally invariant in the three dimensional space $\IR^3$.

\vspace{2cm}


\begin{thebibliography}{30}
\bibitem{foot}
Note that we could have introduced a matrix for the space index as well, as
  ${\cal A}_{\alpha }\equiv B_j^b \,k_{ba} \,e_{ji}$,
  and ${\widetilde { \cal A}}_{\alpha }\equiv
  e^{-1}_{ij}k^{-1}_{ab} \,\(D_j\Phi \)^b$, leading to the self-duality
  equations $B_j^b \,h_{ba} \,g_{ji}=\pm \(D_i\Phi \)^a$, with
  $g=e \,e^T$. We would then introduce a metric in the three
  dimensional space different from the Euclidean one. That is an interesting
  path to pursue but we shall not do it in the present work.
\end{thebibliography}
\end{document}